\documentclass[10pt,twocolumn,letterpaper]{article}

\usepackage{cvpr}

\usepackage{graphicx}
\usepackage{amsmath}
\usepackage{amssymb}
\usepackage{tabularx,booktabs}
\usepackage{adjustbox}
\usepackage[accsupp]{axessibility}

\usepackage{arydshln}

\usepackage[pagebackref,breaklinks,colorlinks]{hyperref}

\usepackage[capitalize]{cleveref}
\crefname{section}{Sec.}{Secs.}
\Crefname{section}{Section}{Sections}
\Crefname{table}{Table}{Tables}
\crefname{table}{Tab.}{Tabs.}

\def\cvprPaperID{895} 
\def\confName{CVPR}
\def\confYear{2023}

\usepackage{hyperref}
\usepackage{url}

\usepackage{graphicx}
\usepackage{multirow}

\usepackage{amsmath,amsfonts,amsthm, amssymb} 
\usepackage{wrapfig}
\usepackage{graphicx}
\usepackage{float}
\usepackage{subcaption}
\usepackage{comment}
\usepackage{enumitem}
\usepackage{cuted}
\usepackage{bm}
\usepackage{sectsty} 
\usepackage{cases}
\usepackage{amssymb}
\usepackage{pifont}

\newcommand{\cmark}{\ding{51}}%
\newcommand{\xmark}{\ding{55}}%

\usepackage{booktabs}
\usepackage{listings}
\usepackage{mathtools}

\usepackage{algorithm}
\usepackage[noend]{algpseudocode}

\usepackage{color}
\usepackage{xcolor}
\usepackage{tabularx}
\usepackage{ulem}

\newcommand{\Erase}{\bgroup\markoverwith{\textcolor{purple}{\rule[.5ex]{2pt}{0.4pt}}}\ULon}

\newtheorem{prop}{Proposition}

\DeclareMathOperator*{\argmax}{arg\,max}
\DeclareMathOperator*{\argmin}{arg\,min}

\newcommand{\beginsupplement}{
  \setcounter{equation}{0}
  \renewcommand{\theequation}{S-\arabic{equation}}
  \setcounter{prop}{0}
  \renewcommand{\theprop}{S-\arabic{prop}}
  \setcounter{table}{0}
  \renewcommand{\thetable}{S-\arabic{table}}
  \setcounter{figure}{0}
  \renewcommand{\thefigure}{S-\arabic{figure}}
  \setcounter{algorithm}{0}
  \renewcommand{\thealgorithm}{S-\arabic{algorithm}}
}

\title{Breaching FedMD: Image Recovery via Paired-Logits Inversion Attack}

\author{
  Hideaki Takahashi\thanks{The University of Tokyo,\newline \texttt{takahashi-hideaki567@g.ecc.u-tokyo.ac.jp}} \and
  Jingjing Liu\thanks{Institute for AI Industry Research, Tsinghua University, \texttt{jjliu@air.tsinghua.edu.cn}} \and
  Yang Liu\thanks{Corresponding Author, Institute for AI Industry Research, Tsinghua University, Shanghai Artificial Intelligence Laboratory, \texttt{liuy03@air.tsinghua.edu.cn}}}

\begin{document}

\maketitle

\begin{abstract}

Federated Learning with Model Distillation (FedMD) is a nascent collaborative learning paradigm, where only output logits of public datasets are transmitted as distilled knowledge, instead of passing on private model parameters that are susceptible to gradient inversion attacks, a known privacy risk in federated learning. In this paper, we found that even though sharing output logits of public datasets is safer than directly sharing gradients,  there still exists a substantial risk of data exposure caused by carefully designed malicious attacks. Our study shows that a malicious server can inject a PLI (Paired-Logits Inversion) attack against FedMD and its variants by training an inversion neural network that exploits the confidence gap between the server and client models. Experiments on multiple facial recognition datasets validate that under FedMD-like schemes, by using paired server-client logits of public datasets only, the malicious server is able to reconstruct private images on all tested benchmarks with a high success rate. 

\end{abstract}

\section{Introduction}

Federated Learning (FL)~\cite{mcmahan2017communication} is a distributed learning paradigm, where each party sends the gradients or parameters of its locally trained model to a centralized server that learns a global model with the aggregated gradients/parameters. While this process allows clients to hide their private datasets, a malicious server can still manage to reconstruct private data (only visible to individual client) from the shared gradients/parameter, exposing serious privacy risks~\cite{zhu2019deep,yin2021see, MOTHUKURI2021619}.

One effective solution against such attacks is Federated Learning with Model Distillation (FedMD)~\cite{li2019fedmd}, where each party uses both its \textit{private data} and a public dataset for local training and sends its predicted logits on the public dataset (termed \textit{public knowledge}) to the server, who then performs aggregation and sends the aggregated logits back to each party for next iteration. Therefore, in FedMD, the server can only access the predicted logits on public dataset, thus preventing leakage from gradients or parameters and retaining data and model heterogeneity with low communication cost~\cite{lyu2020privacy}. Prior work~\cite{lyu2020privacy, 10.1145/3523273, 9411833, momcheva2020sentiment, kumar2021medisecfed} suggests that FedMD is a viable defense to the original FL framework, and many studies~\cite{li2019fedmd, cheng2021fedgems, 10.1145/3523273, 9411833, long2022federated}  consider FedMD as a suitable solution to solving real-life problems. This has catalyzed a broad application of FedMD-like schemes~\cite{chang2019cronus,cheng2021fedgems} in industrial scenarios~\cite{long2020federated, zhao2020privacy, sui2020feded, jiang2022iop, machines10050376, momcheva2020sentiment, kumar2021medisecfed}.

Despite its popularity, few studies have investigated the safety of sharing \textit{public knowledge} in FedMD. 
Contrary to the common belief that FedMD is reliably safe, we found that it is actually vulnerable to serious privacy threats, where an adversary can reconstruct a party's \textit{private data} via shared \textit{public knowledge} only.
Such a reconstruction attack is nontrivial, since to guarantee that the adversary recovers its \textit{private data} strictly from \textit{public knowledge} only, the attack needs to follow two inherent principles: \textit{Knowledge-decoupling} and \textit{Gradient-free}. `Gradient-free' means that the adversary can recover private data without access to gradients. `Knowledge-decoupling' means that the attack `decouples’ private information from learned knowledge on public data only, and directly recovers private data. However, existing inversion attacks~\cite{yang2019adversarial,zhu2019deep,fredrikson2015model} fail to meet the `knowledge-decoupling' requirement as they do not consider the case where the target model is trained on both private and public datasets.
Recent TBI~\cite{yang2019adversarial} proposed a gradient-free method with inversion network, but still relied on availability of private data (as shown in Tab.~\ref{table:comparison}). 

In this paper, we propose a novel \textit{Paired-Logits Inversion} (PLI) attack that is both gradient-free and fully decouples private data from public knowledge. 
We observe that a local model trained on \textit{private data} will produce more confident predictions than a server model that has not seen the \textit{private data}, thus creating a "confidence gap" between a client's logits and the server's logits. Motivated by this, we design a logit-based inversion neural network that exploits this confidence gap between the predicted logits from an auxiliary server model and those from the client model (\textit{paired-logits}).
Specifically, we first train an inversion neural network model based on public dataset. The input of the inversion network is the predicted logits of server-side and client-side models on the public data. 
 The output is the original public data (e.g., raw image pixels in an image classification task). Then, through confidence gap optimization via paired-logits, we can learn an estimation of server and client logits for the target private data. To ensure the image quality of reconstructed data, we also propose a prior estimation algorithm to regulate the inversion network. 
Lastly, we feed those estimated logits to the trained inversion model to generate original private data. 
To the best of our knowledge, this is the first study to investigate the privacy weakness in FedMD-like schemes and to successfully recover private images from shared \textit{public knowledge} with high accuracy. 

We evaluate the proposed PLI attack on facial recognition task, where privacy risks are imperative. Extensive experiments show that despite the fact that logit inversion attacks are more difficult to accomplish than gradient inversion attacks (logits inherently contain less information than gradients), PLI attack  successfully recovers original images in the private datasets across all tested benchmarks. Our contributions are three-fold:
    $1)$ we reveal a previously unknown privacy vulnerability of FedMD; 
    $2)$ we design a novel paired-logits inversion attack that reconstructs private data using only the output logits of public dataset;
    $3)$ we provide a thorough evaluation with quantitative and qualitative analysis to demonstrate successful PLI attacks to FedMD and its variants.

\begin{table}[!ht]
  \caption{
  Comparison between PLI attack and existing inversion attacks. GF: \textit{Gradient-free}. KD: \textit{Knowledge-decoupling.}} 
  \label{table:comparison}
  \centering
  \begin{tabular}{cccc}
    \hline
    \textbf{Method}  & \textbf{Leak} & \textbf{GF} & \textbf{KD} \\
    \hline \hline
    MI-FACE~\cite{fredrikson2015model}  &   logit/gradient      &     \xmark        &        \xmark    \\
    \midrule
    TBI~\cite{yang2019adversarial}  &   logit      &     \cmark        &        \xmark    \\
    \midrule
    \begin{tabular}{c} DLG~\cite{zhu2019deep}, GS~\cite{geiping2020inverting} \\ iDLG~\cite{zhao2020idlg}, CPL~\cite{wei2020framework}\end{tabular} & gradient & \xmark            & \xmark            \\ 
    \midrule
    \begin{tabular}{c} mGAN-AI~\cite{wang2019beyond}, \\ Secret Revealer~\cite{zhang2020secret}, \\ GAN Attack~\cite{hitaj2017deep} \end{tabular} & parameter        &  \xmark           &  \xmark           \\ 
    \midrule
    \textbf{PLI (Ours)} &  logit       &      \textbf{\cmark}       &     \textbf{\cmark}        \\
    \hline
  \end{tabular}
\end{table}

\section{Problem Definition}

\label{subseq:problemdef}

In this study, we investigate the vulnerability of FedMD-like schemes and design successful attacks that can breach FedMD to recover private confidential data. As a case study, we choose image classification task, specifically face recognition, as our focus scenario. This is because in real applications (e.g., financial services, social networks), the leakage of personal images poses severe privacy risks.  
In this section, we first provide a brief overview of FedMD framework with notations. Then, we give a formal definition of image classification task under the federated learning setting.

\textbf{FedMD} ~\cite{li2019fedmd} is a federated learning setting where each of $K$ clients has a small private dataset $D_{k} \coloneqq \{(x^k_{i}, y^k_{i})\}_{i=1}^{N_k}$, which is used to collaboratively train a classification model of $J$ classes. There is also a public dataset $D_{p} \coloneqq \{(x^p_i, y^p_i)\}_{i=1}^{N_p}$ shared by all parties. Each client $k$ trains a local model $f_{k}$ with parameters $W_k$ on either $D_k$, $D_p$ or both, and sends its predicted logits $ \bm l^{k} \coloneqq \{\bm l^{k}_{i}\}_{i=1}^{N_p}$ on $D_p$ to the server, where $ \bm l^{k}_{i}=f_k(W_k; x^{p}_{i})$.
The server aggregates the logits received from all clients to form consensus logits $ \bm l^{p} \coloneqq \{\bm l^{p}_{i}\}_{i=1}^{N_p}$. The consensus logits are then sent back to the clients for further local training (See the complete algorithm in Appendix~B).

Several schemes follow FedMD with slight variations. For example,  FedGEMS~\cite{cheng2021fedgems} proposes a larger server model $f_0$ with parameters $w_0$ to further train on the consensus logits and transfer knowledge back to clients. DS-FL~\cite{itahara2020distillation} uses unlabeled public dataset and entropy reduction aggregation (see Appendix~B for details). These frameworks only use public logits to communicate and transfer knowledge, which is the targeted setting of our attack.   

\begin{figure}[!th]
    \centering
    \includegraphics[width=\linewidth]{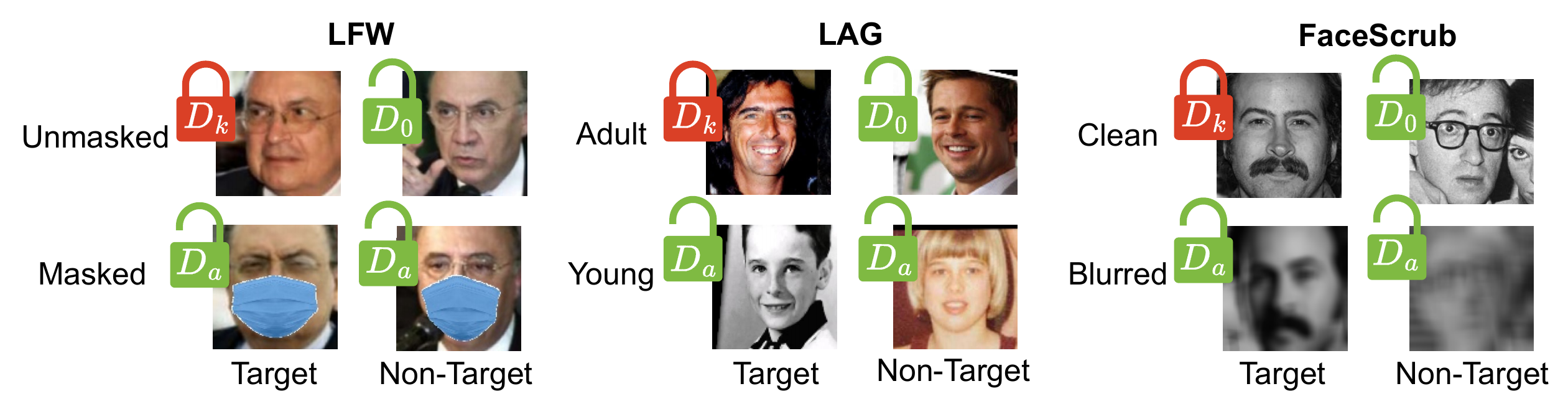}
    \caption{ Examples from three datasets containing private and public data. Public dataset consists of two subsets: $D_p = D_{0} \cup D_{a}$. $D_{0}$ and $D_{k}$ consist of images of non-target and target labels(\textit{Unmasked} (LFW), \textit{Adult} (LAG) and \textit{Clean} (FaceScrub)), while $D_{a}$ contains images from a different domain (\textit{Masked} (LFW), \textit{Young} (LAG), \textit{Blurred} (FaceScrub)).}
    \label{fig:allocation}
\end{figure}

\textbf{Image Classification} task can be defined as follows. Let $(x_i^k$, $y_i^k)$ denote image pixels and class label ID, respectively, and $L$ denote the set of target labels $\bigcup^{K}_{k=1} \{y^{k}_{i}\}^{N_k}_{i=1}$ .We aim to use public logits $\bm l^{k}$ obtained for $D_p$ to reconstruct private class representation $x_j$ for any target label $j \in L$.  We further assume that $D_p$ is made up of two disjoint subsets $D_0 \coloneqq \{(x^0_i, y^0_i)\}$ and $D_a \coloneqq \{(x^a_i, y^a_i)\}$, where $D_0$ consists of public data of non-target labels, while $D_{a}$ contains images from a different domain for all target and non-target labels. For example, in face recognition tasks, the feature space of $D_{a}$ might include an individual's insensitive images, such as masked or blurred faces (see Fig.~\ref{fig:allocation}).  We also suppose the server is honest-but-curious, meaning it cannot observe or manipulate gradients, parameters, or architectures of local models.

\section{Paired Logits Inversion Attack}

In this section, we first describe how to train the inversion neural network for private data recovery (Sec.~\ref{subsec:inversion-neural-network}), then explain how to estimate private logits by optimizing the confidence gap between the logits predictions of server and clients (Sec.~\ref{subsec:confidence gap}). To prevent too much deviation from real data, we also introduce a prior estimation to regulate the inversion network training (Sec.~\ref{subsec:Prior}). Lastly, we will explain how to recover private data using the trained inversion network, the estimated logits and the learned prior (Sec. 3.4).  

\subsection{Inversion Neural Network}
\label{subsec:inversion-neural-network}
Logit-based inversion is a model inversion attack that recovers the representations of original training data by maximizing
the output logits \textit{w.r.t.} the targeted label class of the trained model. 
This inversion typically requires access to model parameters, which is not accessible in FedMD. In order to perform a gradient-free inversion attack, \cite{yang2019adversarial} proposes a training-based inversion method (TBI) that learns a separate inversion model with an auxiliary dataset,  by taking in output logits and returning the original training data. Inspired by this, we insert an inversion attack on the server side of FedMD. However, we show that logit-based inversion is more challenging than gradient-based inversion since logits inherently contain less information about the original data (See Appendix.~E). To tackle the challenge, we train a server-side model $f_0$ on the public dataset with parameters $W_0$. The server-side logits are then updated as :
\begin{equation}
\begin{split}  
     \bm l^{0}_{i}=f_0(W_0;x^{0}_{i}) 
\end{split}
\end{equation} 

Next, we train an inversion model using client-server paired logits, $\bm l^{k}$ and $\bm l^{0}$ on the pubic data subset $D_0$ only, denoted as $G_\theta$: 

\begin{equation}
\begin{split}
    &\min_{\theta} \sum_{i}||G_\theta(p^0_{i,\tau},p^k_{i,\tau})- x^{0}_{i} ||_{2}  \\ 
    &p^{0}_{i,\tau} = \hbox{softmax}(\bm l^0_{i}, \tau), \quad
    p^{k}_{i,\tau} = \hbox{softmax}(\bm l^k_{i}, \tau)
\end{split}
\end{equation} 
where $\hbox{softmax}(,\tau)$ is the softmax function with temperature $\tau$.
Note the distribution gap between $\bm l^{0}$ and $\bm l^{k}$ is the key driver for successfully designing such a paired-logits inversion attack, as will be demonstrated in Sec \ref{subsec:confidence gap}.

\begin{figure}[!th]
    \centering
    \includegraphics[width=\columnwidth]{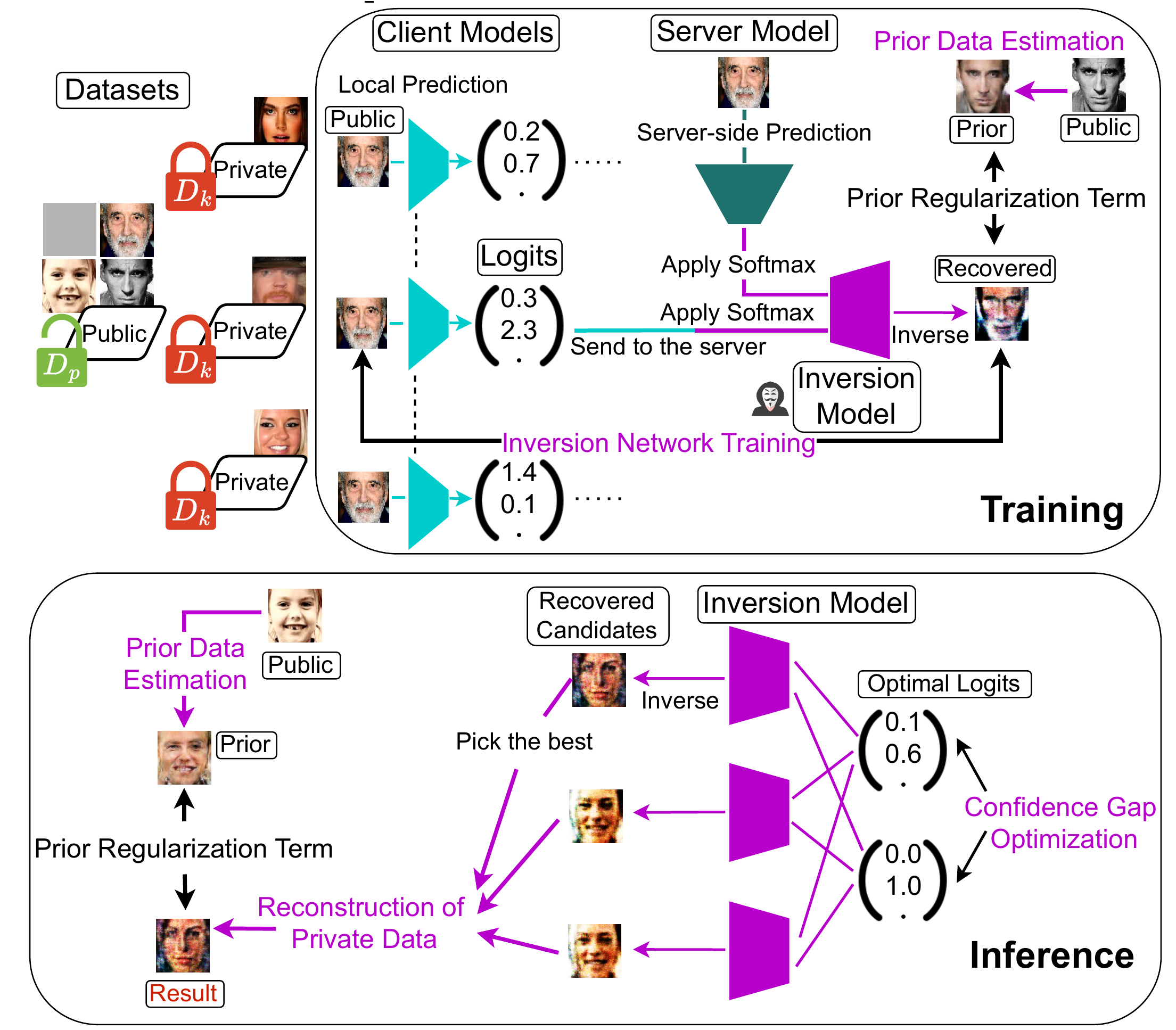}
    \caption{
    Overview of PLI. After receiving the output logits $\bm l^{k}$ on the public data $D_{p}$ from $k$-th client, the server applies softmax to the logits corresponding $D_{0}$ of both client and server models and forwards them to the inversion model $G^{k}_{\theta}$. The inversion model reconstructs the original input. We first optimize the inversion model with Eq.~\ref{eq:loss-bw}, while utilizing the prior data synthesized from $D_{a}$ by transformation model $A_{\phi}$. Then, the server recovers the private date of target labels with optimal logits and trains the inversion model with Eq.~\ref{eq:adjusting}. The final reconstructed class representation will be picked from the set of reconstructed data with Eq.~\ref{eq:choose-distinct}.
    }
    \label{fig:my_label}
\end{figure}

\label{subseq:attack}


\label{subsubsec:training-based-inversion}

To enhance the quality of reconstruction, we leverage the auxiliary domain features $D_a$ to obtain a prior estimation for each data sample $\bar{x}_i$ via a translation algorithm (detailed in Sec.~\ref{subsec:Prior}). The reconstruction objective is therefore summarized as:
\begin{equation}
\label{eq:loss-bw}
  \min_{\theta} \sum_{i}||G_\theta(p^0_{i,\tau},p^k_{i,\tau})- x^{0}_{i} ||_{2} + \gamma || G_\theta(p^0_{i,\tau},p^k_{i,\tau}) - \bar{x}_{i} ||_{2}   \\ 
\end{equation} The second term enforces the recovered data to be close to the prior estimation. 

Since the server already has access to the pubic dataset and its corresponding output logits, it can train $G^{k}_{\theta}$ to minimize Eq.~\ref{eq:loss-bw} via backpropagation (line 6 in Algo~\ref{alg:mia}). 
The architecture of the inversion model typically consists of transposed convolutional layers~\cite{dumoulin2016guide,  yang2019adversarial}. Notice that the inversion model is trained with public data only and did not observe any private data at training time. Once the inversion model is trained, it is able to reconstruct data in the public dataset. 

However, the model cannot be used directly to reconstruct private data with sensitive labels yet, since it has never seen any true logits of the private data, the distribution of which is different from that of public data. Most existing inversion attacks~\cite{fredrikson2015model, zhu2019deep, geiping2020inverting, zhao2020idlg, wei2020framework, wang2019beyond, zhang2020secret, hitaj2017deep} require either gradients or parameters, thus not \textit{gradient-free}. To reconstruct private data without gradients, we propose a new path for estimating the input logits of private data, by exploiting the \textit{confidence gap} between server and clients, as below.

\begin{figure}[!ht]
    \centering
    \includegraphics[width=\linewidth]{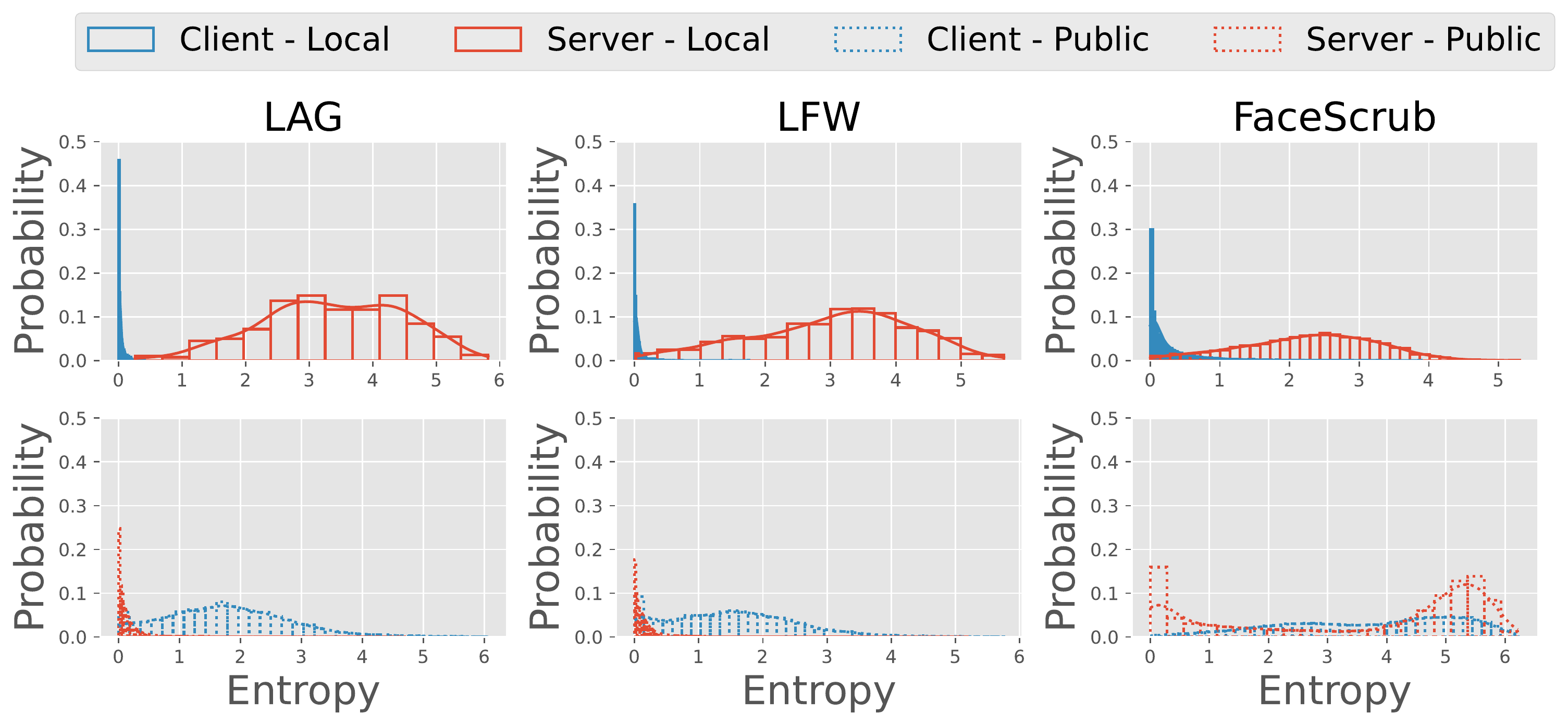}
    \caption{
    Confidence gap between the server and the client under FedMD setting on public and private data. This figure represents the normalized histogram of entropy on public and local datasets and estimated distribution. Lower entropy means that the model is more confident. Client consistently has higher confidence on private dataset than server, indicating a significant confidence gap. }
    \label{fig:confidence}
\end{figure}

\subsection{Confidence Gap Optimization}
\label{subsec:confidence gap}

We observe that the server and client models exhibit a confidence gap on their predictions of the public data, manifested by their predicted logits (Fig.~\ref{fig:confidence}). 
Specifically, the client model is more confident on the private dataset, resulting in a higher entropy of server logits. 
To exploit this, we propose a new metric for optimizing private logits, then analytically obtain the optimal solution, which is used as the input logits for the inversion model to generate private data.

\begin{algorithm}[!t]
\caption{PLI Attack}
\label{alg:mia}
\begin{algorithmic}[1]

\Require The number of communication $T$, clients $K$, the set of target labels $L$, the inversion model $G^{k}_\theta$, the translation model $A_{\phi}$, the global model $f_{0}$, softmax temperature $\tau$, and public dataset $D_{p}$. 

\Ensure Class representations in the private dataset of $L$

\State Generate prior $\bar{x}_{j}$ for each $j \in L$ with $A_{\phi}$.
\Comment{Eq.~\ref{eq:prior}}

\For{$t = 1 \leftarrow T$}
    \State Receive $\{\bm l^{k}_{i}\}$ from $k = 1...K$
    \For {$k = 1 \leftarrow K$}
                \State Train $G^{k}_{\theta}$ with Eq.~\ref{eq:loss-bw} on $ D_{0}$.
            
                \State Train $G^{k}_{\theta}$ with Eq.~\ref{eq:adjusting} for each $j \in L$.
    \EndFor
    \State Update global logits by aggregating $\{\bm l^{k}_{i}\}$
    \State Train $f_{0}$ and send the global logits to each client.
\EndFor

\For{$j \in L$}
    \State Reconstruct $\{\hat{x}^{k}_{j} \leftarrow G^{k}_{\theta}(\hat{p}_{j, \tau}^{0}, \hat{p}_{j, \tau}^{k})\}_{k=1}^{K}$
    \State $\hat{x}_{j} \leftarrow$ Pick the best from $\{\hat{x}^{k}_{j}\}$.
    \Comment {Eq.~\ref{eq:choose-distinct}}
\EndFor
\\
\Return $\{\hat{x}_{j}\}_{j \in L}$

\end{algorithmic}
\end{algorithm}

Specifically, we adopt the following metric to measure the quality of the reconstructed class representation $x^k_{j}$ for arbitrary target label $j$ owned by the $k$-th client: 

\begin{align}
\label{eq:sim}
    Q(x_{j}^{k}) \coloneqq p^{k}_{\underline{j},\tau} + p^{0}_{\underline{j},\tau} + \alpha H(p^{0}_{j,\tau})
\end{align}
where $p^{k}_{\underline{j},\tau}$ and $p^{0}_{\underline{j},\tau}$  denote the $j$-th elements of $p^{k}_{j,\tau}$ and $p^{0}_{j,\tau}$ respectively. 
$H(\cdot)$ is the entropy function, and $\alpha$ is a weighting factor. The first and second terms grow bigger when the client and server are more confident that the recovered data belongs to class $j$, while the third term penalizes the strong confidence of server's prediction, which plays an important role in differentiating public and private feature spaces. With Eq.~\ref{eq:sim}, finding the optimal input logits for our inversion attack against FedMD is equivalent to the following maximization problem:
\begin{align}
\label{eq:max}
    \argmax_{{p^{k}_{j,\tau},p^{0}_{j,\tau}}} Q(x_{j}^{k})
\end{align} Recall that the private and public data domains for the target label $j$ are different. Since the server-side model is not directly trained on the private dataset, we can assume that $f_{0}$ returns less confident outputs on private data compared to $f_{k}$, in this sense $Q$ reinforces the \textit{knowledge-decoupling} principle since $Q$ increases when the reconstructed data of $j$ are similar to the private data, and decreases when too close to the public data. For instance, $Q$ takes a lower value when the reconstructed data is masked, even if its label seems $j$. 

\paragraph{Analytical Solution}
\label{subsubsec:analytical-solution}

We can analytically solve Eq.~\ref{eq:max} \textit{w.r.t.} $p^{k}_{j,\tau}$ and $p^{0}_{j,\tau}$, and the optimal values for target label $j$ that maximizes $Q$, as follows:

\begin{align}
\begin{aligned}
\label{eq:optimal-p}
\hat{p}^{k}_{\underline{u},\tau} &= \begin{cases}
                        1 & (u = j) \\
                        0 & (u \neq j)
                     \end{cases}, \quad
\hat{p}^{0}_{\underline{u},\tau} = \begin{cases}
                        {\frac{\sqrt[\alpha]{e}}{J-1+\sqrt[\alpha]{e}}} & (u = j) \\
                        {\frac{1}{J-1+\sqrt[\alpha]{e}}} & (u \neq j)
                     \end{cases}
\end{aligned}
\end{align} Detailed derivation can be found in Appendix~A. Given an inversion model $G^{k}_{\theta}$ that takes paired logits and returns the original input $x$, we have the following equation:

\begin{equation}
\label{eq:optimal-x}
    \argmax_{x^{k}_{j}} Q(x^{k}_{j}) = G^{k}_{\theta}(\hat{p}^{0}_{j,\tau}, \hat{p}^{k}_{j,\tau})
\end{equation} The empirical impact of feature space gap between public and private data is examined in Appendix~E.

\subsection{Prior Data Estimation} 
\label{subsec:Prior}
To prevent the reconstructed image from being unrealistic, we design a prior estimation component to regulate the inversion network. 
A naive approach is using the average of $D_{0}$ (public data with the same domain) as the prior data 
for any target label $i$:
\begin{equation}
\label{eq:average:prior}
    \bar{x}_{i} = \frac{1}{|D_0|} \sum_{i \in D_0} x^{0}_{i}
\end{equation}
Here the server uses the same prior for all the target labels.

When the public dataset is labeled, such as for FedMD and FedGEMS, the adversary can generate tuned prior for each label by converting $D_a$ with the state-of-the-art translation method such as autoencoder~\cite{chiu2022photowct2, liu2021multiple, qian2019autovc} and GAN~\cite{karras2019style, Kupyn_2019_ICCV, zhu2017unpaired}. Specifically, the server can estimate the prior data $\bar{x}_{i}$ for the target label $i$ with translation model $A_{\phi}$ as follows:
\begin{equation}
\label{eq:prior}
    \bar{x}_{j} = \frac{1}{|D_{a}|} \sum_{i \in D_{a}} A_{\phi}(x^{a}_{i})
\end{equation} 

\subsection{Reconstruction of Private Data}

Once the inversion model is trained with Eq.~\ref{eq:loss-bw}, the attacker uses the optimal logits (obtained from Sec.~\ref{subsec:confidence gap}) to estimate the private data with $G_\theta$. To make the final prediction more realistic, the attacker further fine-tunes the inversion model by restricting the distance between the reconstructed image and their prior data (obtained from Sec.~\ref{subsec:Prior}) in the same way as in Eq.~\ref{eq:loss-bw}:

\begin{equation}
\label{eq:adjusting}
    \min_{\theta} \sum_{j \in L} \gamma || G^{k}_{\theta}(\hat{p}^{0}_{j,\tau}, \hat{p}^{k}_{j,\tau}) -  \bar{x}_{j} ||_{2}
\end{equation}

Finally, given a reconstructed image from each client,  the attacker needs to determine which of the $k$ images $\{\hat{x}^{k}_{j}\}_{k=1}^K$ from $k$ clients is the best estimation for target label $j$. 
We design the attacker to pick the most distinct and cleanest image as the best-reconstructed data based on: $1)$ similarity to groundtruth image via  SSIM metric~\cite{zhou2004ssim}; $2)$ data readability measured by Total Variation (TV)~\cite{rudin1992nonlinear}, which is commonly used as regularizer~\cite{yin2021see, geiping2020inverting}. Specifically, the attacker chooses reconstructed private data $\hat{x}_{j}$ by:

\begin{align}
\label{eq:choose-distinct}
    \argmin_{\hat{x}^{k}_{j}} \sum_{k=1, k \neq k'}^{K} \hbox{SSIM}(\hat{x}^{k}_{j}, \hat{x}^{k'}_{j}) + \beta \hbox{TV}(\hat{x}^{k}_{j})
\end{align} where $\hat{x}^{k}_{j} = G^{k}_{\theta}(\hat{p}^{0}_{j,\tau}, \hat{p}^{k}_{j,\tau})$. Lower SSIM indicates that the image is not similar to any other reconstructed privat pictures, and smaller TV means the image has better visual quality.

\begin{figure}[!th]
    \centering
    \includegraphics[width=\linewidth]{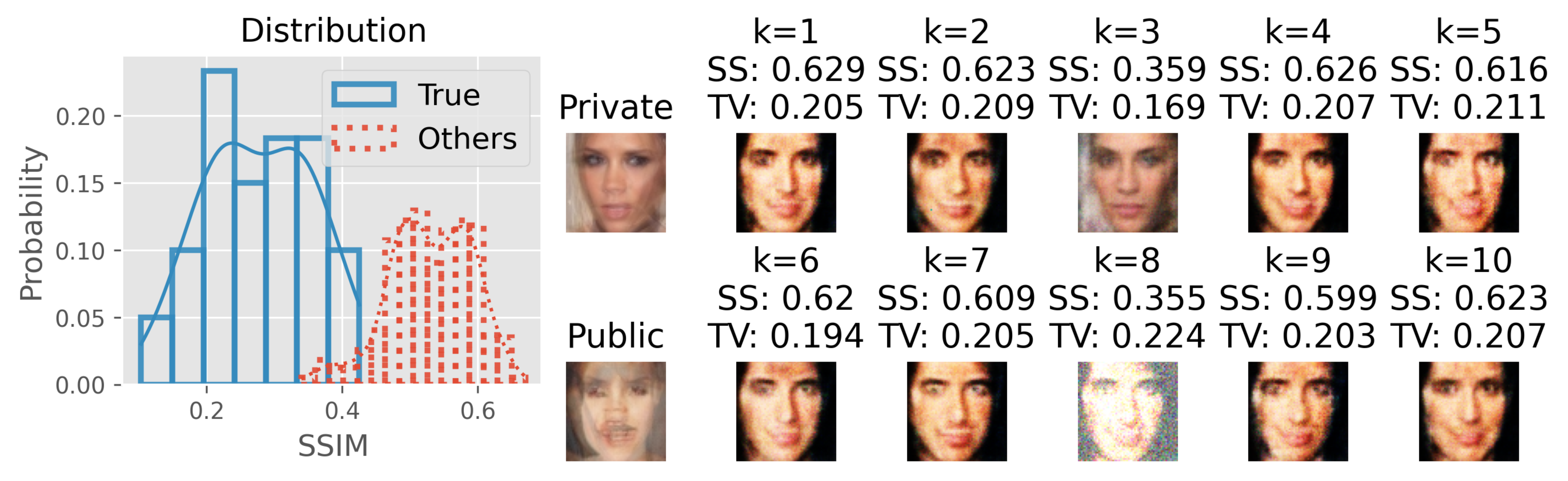}
    \caption{SSIM distribution (left) and Reconstructed Images (right) for LAG.
    }
    \label{fig:rev_eq12}
\end{figure}

The reasoning for choosing Eq.~\ref{eq:choose-distinct} as the criterion is demonstrated in Fig.~\ref{fig:rev_eq12}, where the left figure shows the distribution of the first term of Eq.~12 for clients who own the private data for the target label (blue) and clients who don't (red). Fig.~\ref{fig:rev_eq12} also provides reconstructed examples from all clients, where $k=3$ is the ground-truth client, whose recovered image has the (almost) lowest SSIM. This figure also demonstrates the importance of $\beta$ (the trade-off weight in Eq.~\ref{eq:choose-distinct}, where a highly-noisy image (k=8) results in an even lower SSIM but is penalized by high TV score via a balanced weight ($\beta$ $\geq 0.1$ in this case). 

The complete algorithm is shown in Algo ~\ref{alg:mia}.

\section{Experiments}
\label{seq:experiments}

We evaluate our proposed attack on face recognition task, which inherits high privacy risks in practice. 
We attack three schemes: FedMD~\cite{li2019fedmd}, DS-FL~\cite{itahara2020distillation}, and FedGEMS~\cite{cheng2021fedgems} (see pseudo code in Appendix~B). 
First, we describe detailed comparison between our approach and the baseline method TBI on various attacks. Then, we provide a thorough analysis on the impact of the learned prior, as well as the trade-off between image quality and attack accuracy in different attacks. Ablation studies further provide thorough analysis and insights on the effect of each PLI component and its comparison with gradient-based attack. The source code is available at https://github.com/FLAIR-THU/PairedLogitsInversion

\subsection{Experimental Setup}

\label{subseq:setup}

\paragraph{Datasets and Metrics}

Three standard datasets are used in our experiments: Large-Age-Gap (LAG)~\cite{bianco2017large-age}, Masked LFW ~\cite{phan2021deepface}, and FaceScrub~\cite{7025068}. LAG~\cite{bianco2017large-age} contains both youth and adult images of celebrities. Masked LFW~\cite{phan2021deepface} is created by automatically adding masks to the faces of images in the LFW dataset~\cite{LFWTechUpdate}. For LAG and LFW, we treat adult and non-masked images as sensitive features, and young and masked images as insensitive features. We use the images of 1000 subjects who have the highest numbers of photos, so the number of classes is 1000. Among these, we randomly pick 200 classes as private labels, and treat the remaining classes as public labels. For FaceScrub~\cite{7025068}, we randomly select 330 persons and treat all their pictures as the public dataset. Then, we blur half of the images of the remaining 200 individuals with box kernel and merge them to the public dataset as auxiliary domain $D_a$. The private datasets consist of the remaining clean photos. All pictures of each dataset are resized to 64x64 and normalized to [-1, 1].

An attack is considered successful when: $\hbox{SSIM}(\hat{x}_j, x_j) > max(\hbox{SSIM}(\hat{x}_j, x_{-j}), \hbox{SSIM}(\hat{x}_j, x_p))$. $x_j$: averaged private image of label $j$. $x_{-j}$: averaged private image of each label except $j$. $x_p$: averaged public image of any label. Assume \textit{M} is the number of target labels with successful attack, and \textit{N} is the number of all target labels. \textit{AttackAccuracy} is defined by $M/N$. 

\paragraph{Attack Targets}

We target three types of schemes: FedMD~\cite{li2019fedmd}, DS-FL~\cite{itahara2020distillation}, and FedGEMS~\cite{cheng2021fedgems}. For simplicity, the server and clients use the same neural network with a convolutional layer, a max-pooling layer, and a linear layer with ReLU (Code.~1 in Appendix~B). DS-FL assumes that the public dataset is unlabeled, so the total number of classes in DS-FL equals the number of classes in the local dataset, 200. Classification loss is cross entropy for all schemes, and distillation loss is L1-loss for FedMD, KL divergence for FedGEMS, and cross-entropy for DS-FL (as in original papers). See Appendix~C for the detailed hyper parameters.

For the translation model $A_{\phi}$ for learned prior, the attacker trains CycleGAN~\cite{zhu2017unpaired} for LAG and LFW, and DeblurGAN-v2~\cite{Kupyn_2019_ICCV} for FaceScrub in advance on the public dataset when it is labeled (Eq.~\ref{eq:prior}). CycleGan is used for facial aging and mask removing in some work~\cite{sharma2022comparative, farahanipad2022gan, Palsson_2018_CVPR_Workshops}, and DeblurGAN-v2 is one of the state-of-the-art methods to deblur images. For DS-FL, the server uses the average of public sensitive features as the prior (Eq.~\ref{eq:average:prior}). For the inversion model, we adopt the same architecture used in TBI (see Code.~2 in Appendix B). Detailed hyper-parameters can be found in Appendix C.

\subsection{Comparison with Baseline}
\label{subseq:result}

\begin{table*}[!ht]
    \begin{tabular}{c|ccc|ccc|ccc}
    \toprule
    Dataset & \multicolumn{3}{|c|}{\textbf{FaceScrub}} & \multicolumn{3}{|c|}{\textbf{LAG}} & \multicolumn{3}{|c}{\textbf{LFW}} \\
    \midrule
    Scheme &      DS-FL & FedGEMS & FedMD &  DS-FL & FedGEMS & FedMD &  DS-FL & FedGEMS & FedMD \\
    \midrule
     TBI ($K$ = 1)    &      87.0 &     1.0 &  92.5 &  70.0 &     0.5 &  16.5 &  73.5 &     2.5 &   2.0 \\
     PLI ($K$ = 1)   &      \textbf{91.5} &    \textbf{29.0} &  \textbf{94.0} &  \textbf{71.0} & \textbf{17.0} & \textbf{60.0} & \textbf{99.5} & \textbf{91.0} &  \textbf{99.5} \\
     \midrule
     TBI ($K$ = 10)   &       2.0 &     0.5 &   7.0 &   6.5 &     0.0 &   0.0 &  17.5 &     9.5 &  10.0 \\
     PLI ($K$ = 10)  &  \textbf{62.5} &  \textbf{20.0} & \textbf{74.5} & \textbf{15.0} &    \textbf{26.5} &  \textbf{63.5} &  \textbf{15.5} &  \textbf{71.5} & \textbf{79.0} \\
    \bottomrule
    \end{tabular}
    \caption{Results on attack accuracy (\%).
    }
    \label{tab:results:main}
\end{table*}

\begin{figure}[!ht]
    \centering
    \includegraphics[width=\linewidth]{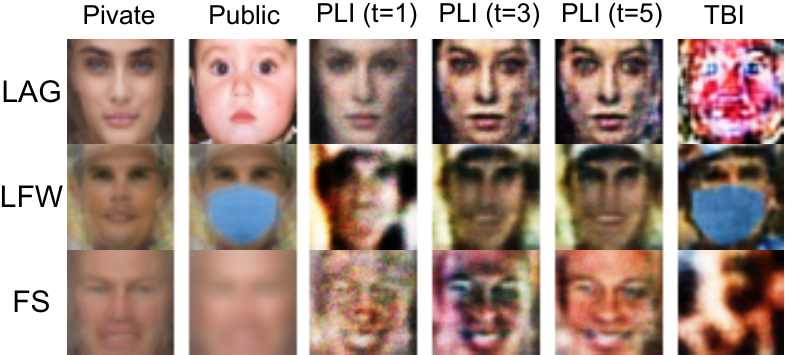}
    \caption{Examples of reconstructed images during training by our method where $t$ represents the number of communications (FS: FaceScrub). "TBI" represents the reconstructed results from TBI. Although TBI generates insensitive images, our PLI successfully reconstructs sensitive images as the learning progresses.}
    \label{fig:results:main}
\end{figure}

Tab.~\ref{tab:results:main} reports the attack accuracy of PLI in comparison with TBI. We use $\tau$=3.0 and $\gamma$=0.03 here as the empirically optimal setting across schemes and datasets.
Results show that PLI outperforms TBI in all settings. Fig.~\ref{fig:results:main} shows some reconstructed images from FedMD. The malicious party tries to recover private images. As the communication goes (higher $t$), the quality of reconstructed images improves (higher similarity to the grounthuth of private images). These images successfully capture the distinct sensitive features of the original private data, demonstrating that public logits can leak private information. On the other hand, the reconstructed images from TBI tend to be closer to the insensitive public features. More recovered examples from FedGEMS and DS-FL can be found in Appendix~D. 

In general, attack accuracy on FedGEMS is lower than that on the other two schemes. 
We hypothesize that this is because FedGEMS trains iteratively on both private and hard-labeled public data, compared to FedMD that trains each model on labeled public dataset only. DS-FL does not even use labels on public data. Thus, local models in FedGEMS are more affected by the public dataset compared to FedMD and DS-FL, and more difficult to accurately recover private data.

\subsection{Analysis on Impact of Prior }

We also discuss how the quality of
prior data is related to optimal $\gamma$, the tunable weight for prior.

\begin{figure}[!ht]
    \centering
    \includegraphics[width=\linewidth]{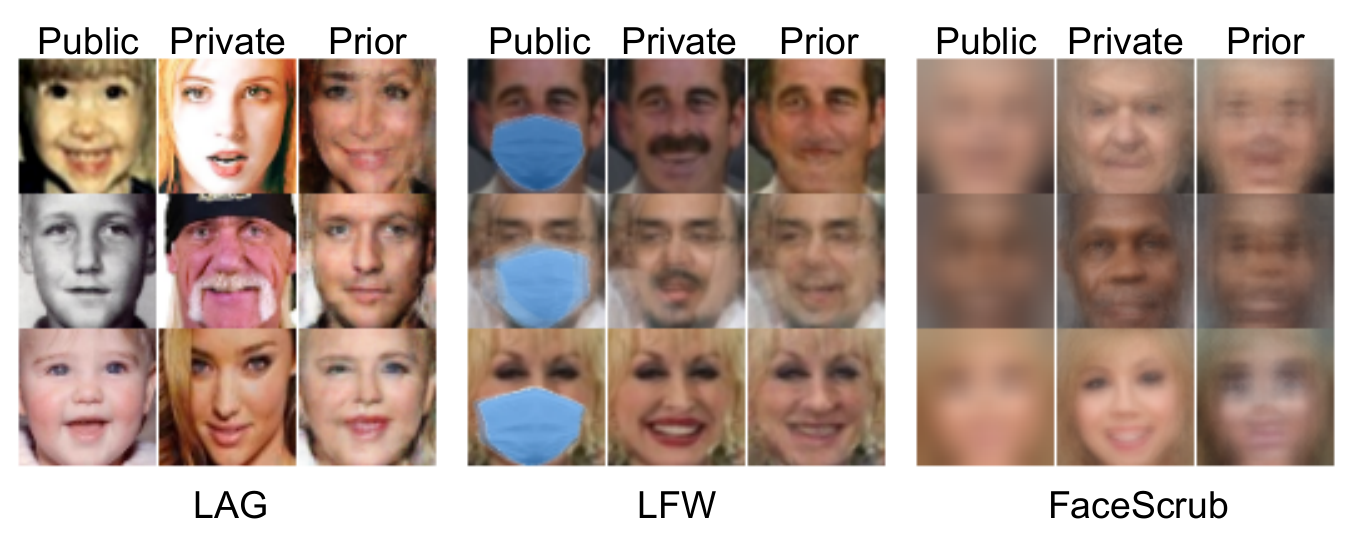}
    \caption{Examples of Prior for the labeld public dataset. While we can obtain relatively good prior images for LFW, the prior pictures for LAG fail to estimate the change of the shape of faces. For FaceScrub, the prior images are still not sharp enough.}
    \label{fig:prior}
\end{figure} 

Fig.~\ref{fig:prior} shows some qualitative examples of prior synthesized from the generator $A_\phi$ with the labeled public data. For LFW, CycleGAN successfully replaces the masked region with unmasked mouth. However, some distinct features under the mask, such as beard or lip color, are still not recovered. For LAG, faces in prior data look older than the public images, but are still more similar to private ones. Specifically, CycleGAN renders public images in adult appearances while keeping the original face outline. In addition, the recovered image is often too young or too old compared to the private image, since the attacker does not know the true age of the target person.
For FaceScrub, DeblurGAN-v2 can deblur public images to some extent, but estimated prior data
is still closer to public images.

\begin{figure}[!th]
    \centering
    \includegraphics[width=\linewidth]{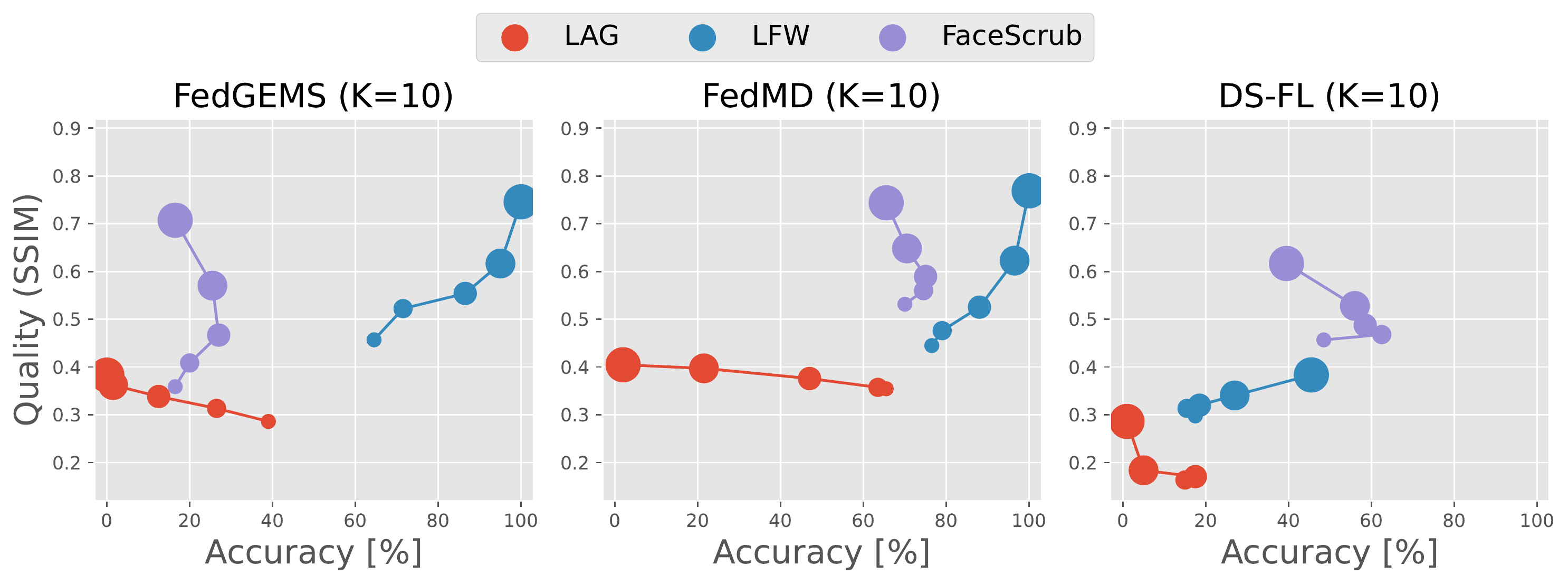}
    \caption{Trade-off between quality and accuracy with various $\gamma$. Dot size corresponds to the magnitude of $\gamma$. Larger $\gamma$ leads to high quality but ruins attack accuracy for LAG and FaceScrub, where the prior images are not close enough to the ground-truth private images. For LFW, the prior sufficiently resemble the private images, and increasing $\gamma$ improves both accuracy and quality. 
    }
    \label{fig:tradeoff:gamma}
\end{figure}

\begin{figure}[!th]
    \centering
    \includegraphics[height=3.7cm]{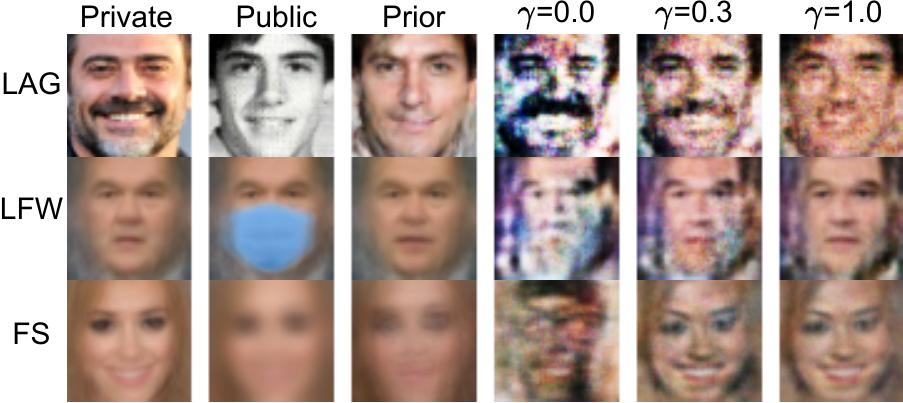}
    \caption{Effect of $\gamma$ in FedMD (FS: FaceScrub) with fixed $\tau$. Higher $\gamma$ makes the reconstructed images noiseless, but sometimes drops distinctive elements (e.g., beard in the first row). 
    }
    \label{fig:example:gamma}
\end{figure}

Fig.~\ref{fig:tradeoff:gamma} reports attack accuracy and SSIM value with various weights of prior $\gamma$ with fixed $\tau$=3.0. Note that the attacker cannot access GAN-based prior for DSFL, thus uses the average of all sensitive public features as prior (Eq.~\ref{eq:average:prior}). While PLI can reconstruct private class representations without prior($\gamma$=0.0), using prior as the regularization improves the attack performance in some cases. On the one hand, when the prior is relatively reliable (such as for LFW), higher $\gamma$ increases both attack accuracy and SSIM. On the other hand, for LAG and FaceScrub, lower $\gamma$ increases attack accuracy but damages quality. This effect of $\gamma$ is reasonable as increasing $\gamma$ makes the reconstructed images closer to prior images, which generally decreases the noise but also eliminates some distinctive features (such as beards) that the prior image does not have (see Fig.~\ref{fig:example:gamma}).

\subsection{Attack Accuracy vs. Image Quality}

Experiments show that temperature $\tau$ controls the trade-off between attack accuracy and image quality. Fig.~\ref{fig:tradeoff:tau} shows the attack performance on various temperature $\tau$ with fixed $\gamma=0.1$ (see Appendix~D for the detailed numerical values).
We observe that higher $\tau$ improves accuracy but damages quality. The influence of $\tau$ is consistent with previous work~\cite{itahara2020distillation}, suggesting that a higher $\tau$ amplifies non-highest scores of predicted logits 
while losing class information. We also find that the best temperature for DS-FL is lower in some cases. It is natural because the predicted logit on public data is more vague when the trained model has not observed actual public labels. Fig.~\ref{fig:example:tau} shows some qualitative examples, and we can observe that increasing $\tau$ leads to reconstructed images being created by compositing multiple classes, which preserves distinctive elements but reduces the quality of image. The same pattern occurs with TBI (more results can be found in Appendix~D).

\begin{figure}
    \centering
    \includegraphics[width=\linewidth]{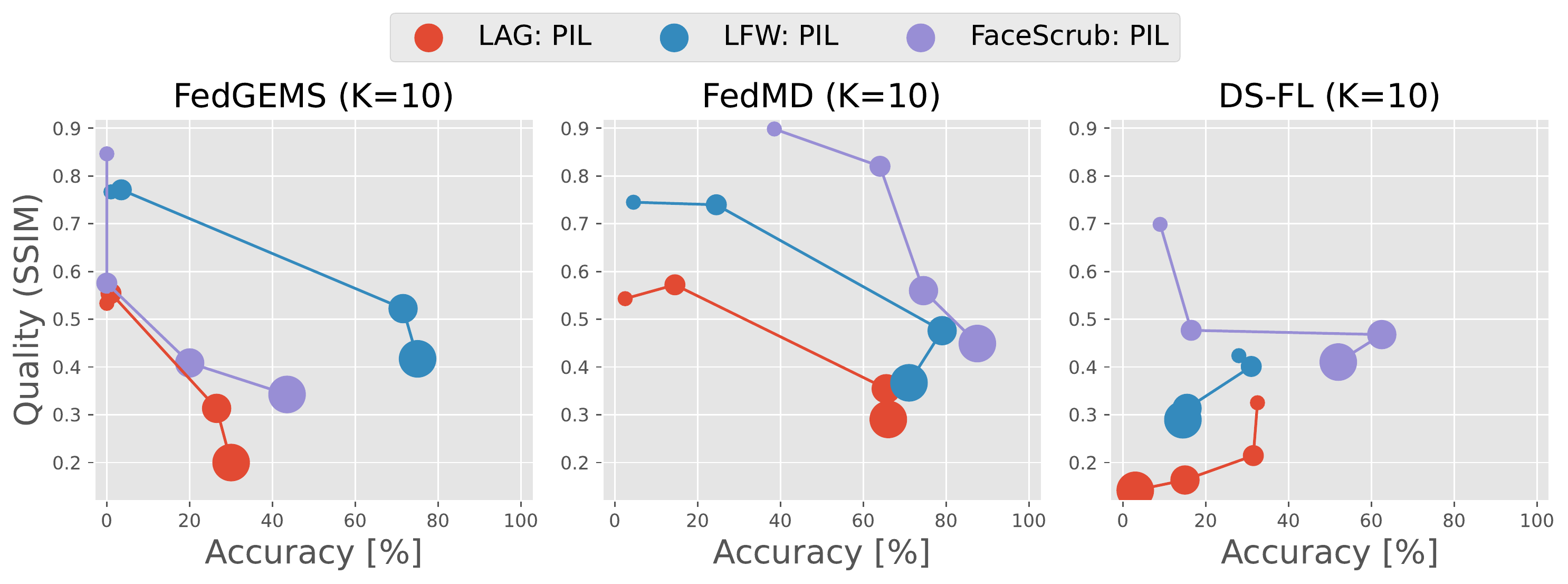}
    \caption{Trade-off between quality and accuracy with various $\tau$. Dot size is proportional to the magnitude of $\tau$. Larger $\tau$ leads to high accuracy but ruins quality for FedMD and FedGEMS, where the local models can use the labeled public dataset. The appropriate $\tau$ is lower for DSFL, where the public data is unlabeled and the predicted logits on the public data are more ambiguous.
    }
    \label{fig:tradeoff:tau}
\end{figure}

\begin{figure}
    \centering
    \includegraphics[height=3.7cm]{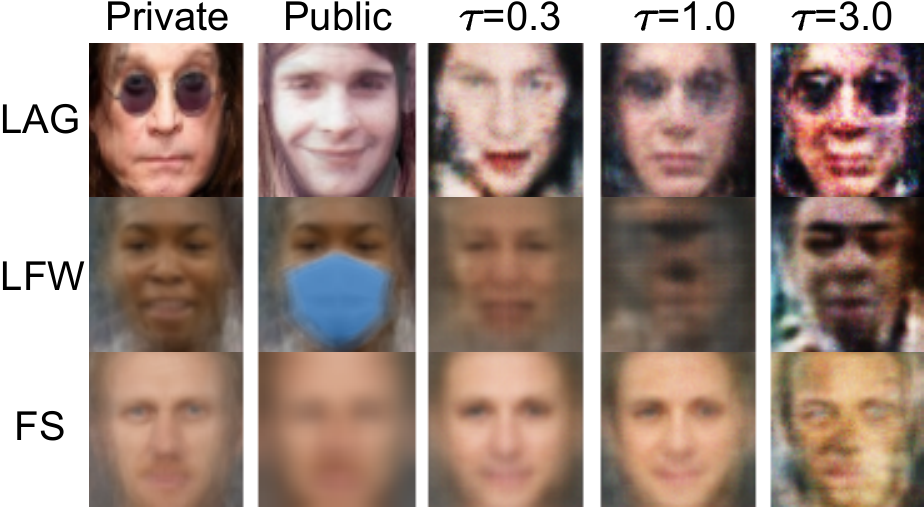}
    \caption{Effect of $\tau$ in FedMD with fixed $\gamma$ (FS: FaceScrub).
    Higher $\tau$ preserves the characteristic features but damages readability (e.g., skin color in the second row). 
    } 
    \label{fig:example:tau}
\end{figure}

In summary, these experiments exhibit a trade-off between accuracy and quality, as optimizing attack accuracy does not guarantee improved quality. By finding a sweet spot (via $\gamma$ and $\tau$), we can empirically learn an optimal setting with high accuracy and acceptable quality.

\subsection{Ablation Study}

\begin{table}[!th]
    \centering
    \begin{tabular}{c|ccc}
    \toprule
     &  \textbf{FaceScrub} &     \textbf{LAG} &     \textbf{LFW} \\
    \hline
    $p^{k}_{j}$         &   55.5 &  57.0 &  71.5 \\
    $+ p^{0}_{j} + \alpha H(p^{0})$ &   \textbf{70.0} &  \textbf{65.5} &  \textbf{76.5} \\
    \bottomrule
    \end{tabular}
    \caption{Ablation studies on $Q$ (FedMD with $K$=10) in terms of attack accuracy. PLI (second row) outperforms using only client-side logits $p^{k}_{j}$ in terms of  accuracy.} 
    \label{tab:ablation:q}
\end{table}

\begin{figure}[!th]
    \centering
    \includegraphics[height=3.7cm]{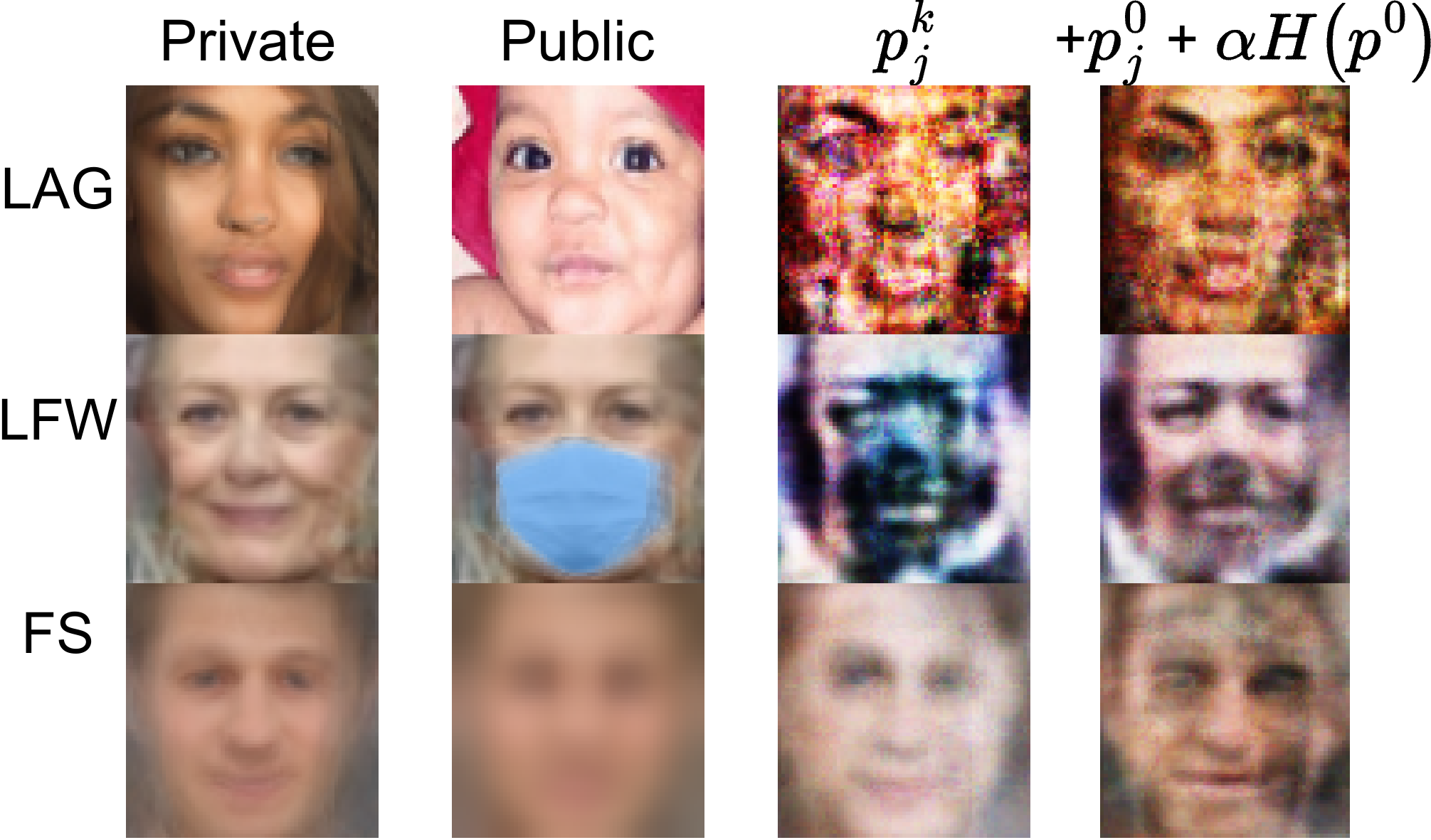}
    \caption{Qualitative analysis: eliminating $\alpha H(p^{0})$ and $p^{0}_{j}$ from $Q$ leads to reconstructed images with featureless faces or higher noise. (FS: FaceScrub)
    }
    \label{fig:ablation:examples}
\end{figure}

We further investigate the influence of each term of $Q$. Tab.~\ref{tab:ablation:q} shows the results of ablation studies on FedMD of $K=10$, $\tau=3.0$ and $\gamma=0$, by gradually adding server and local logits of Eq.~\ref{eq:sim}. Note that the second row is PLI, and the optimal logits for the first row is $(\hat{p}^{k}_{j}, \hat{p}^{0}_{j}) = (1, -)$. The results show that it is effective to add both $p^{0}_{j}$ and $\alpha H(p^{0})$. This phenomenon validates the importance of taking into consideration the confidence gap between the server and client logits. The effects of public dataset size and magnitude of feature domain gap are reported in Appendix~D. Appendix~E and F also show that gradients lead to more severe privacy violations than public logits.

\vskip\baselineskip

\paragraph{Limitations }
One limitation of our method is that it generates representative data of each class, not pixel-wise accurate image within the private dataset. Since the attacker cannot obtain anything directly calculated from the private dataset, reconstructing the exact image is still challenging.

\section{Related Work}

Federated Learning~\cite{mcmahan2017communication} is one of the representative algorithms for distributed learning. To overcome data and model heterogeneity, many studies focus on knowledge distillation~\cite{lyu2020privacy}. In addition to the three schemes targeted in this work, there are several methods~\cite{lin2020ensemble, wu2021fedkd} that communicate both distilled knowledge and gradients. While these schemes are the combination of gradient-based and knowledge-distillation-based federated learning, they are subject to existing gradient-based attacks, which are more severe attacks (as shown in Appendix E and F).

Model inversion attack aims to recover training data via machine learning models, which takes in information about the model (output logits, gradients, or parameters) and generates training data. 
Gradient inversion attack 
reconstructs the original training data by minimizing the distance between the received gradient that each client calculates on its local dataset and model 
and the gradient of synthesized data on the loss function, with model parameters used on the client side
~\cite{zhu2019deep, geiping2020inverting, zhao2020idlg, wei2020framework}. Parameter-based inversion attacks utilize parameters of the target model. For example,~\cite{wang2019beyond, zhang2020secret, hitaj2017deep} build GAN with built-in local models to recover the private data. Such attacks may not apply to FedMD scenarios when the server cannot access the parameters of client models. 
MI-FACE~\cite{fredrikson2015model} reconstructs training data by directly optimizing synthesized data, so the learned model generates a high probability for the target class on that data. However, it requires the gradient of the model w.r.t synthesized data, thus not satisfying the gradient-free principle. In this sense, MI-FACE can be categorized as a combination of gradient-based and logit-based attacks. 

\section{Conclusion}
\label{seq:conclusion}

This paper introduces an attack against FedMD and its variants that reconstructs private training data from the output logits of the model trained on both private and public datasets. Our proposed attack does not rely on gradients, parameters or model structures. Experiments on face recognition benchmarks demonstrate that confidential information can be recovered from output logits of public datasets. By demonstrating FedMD-like methods are not as safe as previously believed, we hope to inspire future design of more secure FL algorithms for diverse tasks. Potential future work includes adaptively choosing $\gamma$ based on the quality of the prior, and investigating protection mechanisms such as differential privacy~\cite{abadi2016deep, dwork2014algorithmic} and homomorphic encryption~\cite{aono2017privacy, 9452005}.

\section{Acknowledgement}
\label{seq:ack}
This work was supported by the National Key R\&D Program of China under Grant No.2022ZD0160504, Tsinghua-Toyota Joint Research Institute inter-disciplinary Program, and Xiaomi AI Innovation Research under grant No.202-422-002.
{\small
\bibliographystyle{ieee_fullname}
\bibliography{ref}
}

\beginsupplement

\appendix

\section{Deviation of Eq.~5}


\begin{proof}

We solve Eq.~5 independently for $p_{j, \tau}^{k}$ and $p_{j, \tau}^{0}$. The optimal $p^{k}_{j, \tau}$ is obviously as follows;

\begin{equation}
    \hat{p}^{k}_{\underline{u}, \tau} = \begin{cases}
                            1 & (u = j) \\
                            0 & (u \neq j)
                         \end{cases}
\end{equation} Next, we solve the optimal $p_{j, \tau}^{0}$ under the constraint of its sum equal to one. 

\begin{align}
    \label{eq:max-ps}
    \max_{p_{j, \tau}^{0}} p^{0}_{\underline{j}, \tau} + \alpha H(p_{j, \tau}^{0}) \\
    \label{eq:st}
    s.t. \sum_{u=1}^{J} p^{0}_{\underline{u}, \tau} = 1
\end{align} Substituting Eq.~\ref{eq:st}, we can arrange Eq.~\ref{eq:max-ps} as follows;

\begin{align}
    \max_{p_{j, \tau}^{0}} \quad &1 - \sum_{u=1, u \neq j}^{J} p^{0}_{\underline{u}, \tau}
    - \alpha (\sum_{u=1, u \neq j}^{J} p^{0}_{\underline{u}, \tau} \log p^{0}_{\underline{u}, \tau}) \\
    &- \alpha (1 - \sum_{u=1, u\neq j}^{J} p^{0}_{\underline{u}, \tau}) \log (1 - \sum_{u=1, u\neq j}^{J} p^{0}_{\underline{u}, \tau})
\end{align} , which requires the bellow for all $u \neq j$:

\begin{equation}
    -1 + \alpha (\log p^{0}_{\underline{j}, \tau} + 1) - \alpha (\log p^{0}_{\underline{u}, \tau} + 1) = 0
\end{equation} Then, we have that

\begin{equation}
    \label{eq:ratio}
    \forall u \in \{u: 1 \leq u \leq J, u \neq j \}, \quad \sqrt[\alpha]{e} = \frac{p^{0}_{\underline{j}, \tau}}{p^{0}_{\underline{u}, \tau}}
\end{equation} Eq.~\ref{eq:st} and~\ref{eq:ratio} lead to

\begin{equation*}
    \hat{p}^{0}_{\underline{u}, \tau} = \begin{cases}
                        {\frac{\sqrt[\alpha]{e}}{J-1+\sqrt[\alpha]{e}}} & (u = j) \\
                        {\frac{1}{J-1+\sqrt[\alpha]{e}}} & (u \neq j)
                     \end{cases}
\end{equation*}
\end{proof}

\section{Protocols \& Architectures}
\label{appendix:protocols}

\begin{algorithm}[!th]
\caption{FedMD 
}
\label{alg:fedmd}
\begin{algorithmic}[1]
\Require Private datasets $\{D_{k}\}^{C}_{k=1}$, public dataset $D_{p}$, local models $\{f_{k}\}^{C}_{k=1}$, global model $f_{0}$, number of communications $T$.
\State Each client trains $f_{k}$ on $D_{p}$
\State Each client trains $f_{k}$ on $D_{k}$

\For{$t = 1 \leftarrow T$}
    \State Each client sends the set of public logits $\{\bm l^{k}_{i}\}$ 
    \State The server computes the global logits:
    \State $\quad \quad {\bm l_{p}} = \frac{1}{K} \sum_{k=1}^{K} \bm l^{k}$
    \State Each client receives $\bm l_{p}$ and trains $f_{k}$ on $\{D_{p}, \bm l_{p}\}$
    \State Each client trains  $f_{k}$ on $D_{k}$
    \State The server trains  $f_{0}$ on $D_{p}$
\EndFor
\end{algorithmic}
\end{algorithm}

\begin{algorithm}[!th]
\caption{FedGEMS}
\label{alg:fedgems}
\begin{algorithmic}[1]
\Require Private datasets $\{D_{k}\}^{K}_{k=1}$, public dataset $D_{p}$, local models $\{f_{k}\}^{K}_{k=1}$, global model $f_{0}$, number of communications $T$.
\For{$t = 1 \leftarrow T$}
    \State The server selectively trains $f_{0}$ on $\{D_{p}, {\bm l_{p}}, \bm l_{k}\}$
    \State The server computes the global logits:
    \State $\quad \quad {\bm l_i^{p}} = f_{0}(W_p;x_i^{p})$
    \State Each client trains $f_{k}$ on $\{D_{p}, {\bm l_{p}}\}$
    \State Each client trains $f_{k}$ on $D_{k}$    
    \State Each client sends the set of public logits $\{\bm l^{k}_{i}\}$ 
\EndFor
\end{algorithmic}
\end{algorithm}

\begin{algorithm}[!th]
\caption{DS-FL}
\label{alg:DS-FL}
\begin{algorithmic}[1]
\Require Private datasets $\{D_{k}\}^{K}_{k=1}$, public dataset $D_{p}$, local models $\{f_{k}\}^{K}_{k=1}$, global model $f_{0}$, number of communications $T$.
\For{$t = 1 \leftarrow T$}
    \State Each client trains $f_{k}$ on $\{D_{k}\}$ 
    \State Each client sends the set of public logits $\{\bm l^{k}_{i}\}$ 
    \State The server computes the global logits:
    \State $\quad \quad {\bm l_{p}} = \hbox{ERA}(\sum_{k=1}^{K} \frac{\bm l^{k}}{K})$
    \State Each client trains $f_{k}$ on $\{D_{p}, {\bm l_{p}}\}$
    \State The server trains $f_{0}$ on $\{D_{p}, {\bm l_{p}}\}$
\EndFor
\end{algorithmic}
\end{algorithm}

Alg.~\ref{alg:fedmd},~\ref{alg:fedgems}, and~\ref{alg:DS-FL} are the pseudo-codes of each protocol, where we additionally train the global model on the public dataset at line 9 in FedMD. Code.~\ref{lst:base-model} and~\ref{lst:inv-model} are the implementation of global, local, and inversion models.

\begin{lstlisting}[language=Python, caption=Server and local models, label={lst:base-model}]
nn.Sequential(
    nn.Conv2d(3, 32, kernel_size=(3, 3),
    stride=1, padding=0),
    nn.ReLU(),
    nn.MaxPool2d(kernel_size=(3, 3),
    stride=None, padding=0),
    nn.Flatten(),
    nn.Linear(12800, output_dim))
\end{lstlisting}

\begin{lstlisting}[language=Python, caption=Inversion model, label={lst:inv-model}]
nn.Sequential(
    nn.ConvTranspose2d(input_dim, 1024,
    (4, 4), stride=(1, 1)),
    nn.BatchNorm2d(1024),
    nn.Tanh(),
    nn.ConvTranspose2d(1024, 512, (4, 4), 
    stride=(2, 2), padding=(1, 1)),
    nn.BatchNorm2d(512),
    nn.Tanh(),
    nn.ConvTranspose2d(512, 256, (4, 4),
    stride=(2, 2), padding=(1, 1)),
    nn.BatchNorm2d(256),
    nn.Tanh(),
    nn.ConvTranspose2d(256, 128, (4, 4),
    stride=(2, 2), padding=(1, 1)),
    nn.BatchNorm2d(128),
    nn.Tanh(),
    nn.ConvTranspose2d(128, 3, (4, 4),
    stride=(2, 2), padding=(1, 1)),
    nn.Tanh())
\end{lstlisting}

\section{Hyper Parameters}
\label{appendix:sec:hyper-params}

For FedMD, the number of consensuses, revisit, and server-side epochs are 1, and the number of transfer epochs is 5. For FedGEMS, the client-side epoch on both public and private datasets are 2, and the number of server-side epochs is 1. For DS-FL, the number of epochs of local update and distillation is 2, and the number of epochs of server-side distillation is 1. Thus, local models iterate both datasets ten times, and the server iterates the public dataset 5 times in all settings.

We use Adam optimizer with a learning rate of 1e-3 and batch size of 64. The number of clients is 1 or 10. Following the original papers, we set the parameter of FedGEMS $\epsilon$ to 0.75 and the temperature of DS-FL to 0.1. Although the original FedMD does not use a server-side model, we train a server-side model on the labeled public dataset. The number of communication is 5 in all schemes. 

For PLI, the attacker trains the same architecture (Code.~2 in Appendix~B) used in the original TBI as $G_{\theta}$, with Adam optimizer whose learning rate is 3e-5, weight-decay is 1e-4, and batch size is 8. We experiment with 0.3, 1, 3, and 5 for temperature $\tau$, 0.0, 0.03, 0.1, 0.3, and 1.0 for $\gamma$, 5.0 for $\alpha$, and 0.1 for $\beta$. The number of epochs $M$ in each communication is 3. For CycleGAN and DeblurGAN-v2, We set a learning rate of 1e-4 and 2e-4 for each, with a batch size of 1 and 100 epochs.

For comparison, we attack the victim with TBI with the same model architecture, optimizer, and data augmentation. As in the original paper, TBI trains a single inversion model with all available logits on the public dataset. We also apply gradient inversion attacks to FedAVG, the standard scheme of FL, as the baseline. Same as other FedMD-like schemes, the number of communication is 3, and the epoch of local training is 2, but we do not use the public dataset in FedAVG (See Appendix~F for details).

\section{Additional Results}
\label{appendix:seq:additional_results}

\paragraph{Impact of $\gamma$ and feature space gap}

\begin{figure*}
    \centering
    \includegraphics[height=8.5cm]{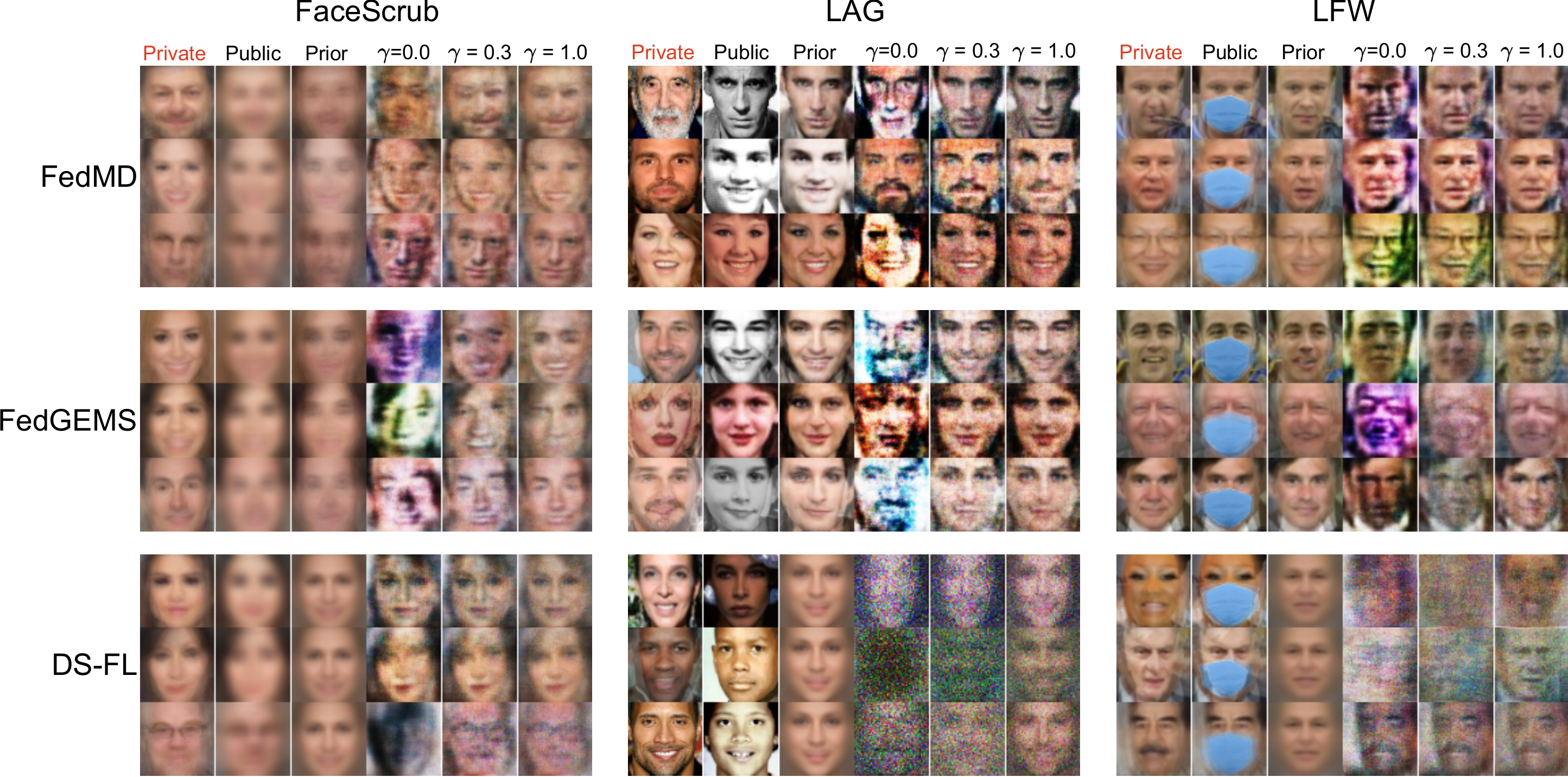}
    \caption{Example of reconstructed images with various $\gamma$. Higher $\gamma$ makes the recovered images closer to the prior images. Note that the attacker use the average of the public sensitive images as the prior for the unlabeled public dataset of DS-FL.}
    \label{fig:examples:gamma}
\end{figure*}

As discussed in Sec.~3.3, the better the quality of prior images is, the higher effective $\gamma$ is. Tab.~\ref{tab:result:gamma:accuracy} and Tab.~\ref{tab:result:gamma:ssim} show the results of attack accuracy and SSIM with different $\gamma$, which are visualized in Fig.~7. The recovered images with different $\gamma$ can be found in Fig.~\ref{fig:examples:gamma}. Tab.~\ref{tab:prior:ssim} also reports the SSIM between the GAN-based prior images based on the labeled public dataset and the private images. It is natural to assume that this SSIM correlates to the feature space gap since the smaller feature space gap between the sensitive and insensitive data should improve the quality of prior data. Thus, these tables validate the proportional relationship between the feature space gap and the optimal $\gamma$ for attack accuracy.

\begin{table} 
\begin{adjustbox}{width=\columnwidth,center}
    \begin{tabular}{c|ccc|ccc|ccc}
    \toprule
    Dataset & \multicolumn{3}{|c|}{FaceScrub} & \multicolumn{3}{|c|}{LAG} & \multicolumn{3}{|c}{LFW} \\
    \midrule
    Scheme &                  DS-FL & FedGEMS &  FedMD &  DS-FL & FedGEMS &  FedMD &  DS-FL & FedGEMS &  FedMD \\
    \midrule
    $\gamma$ = 0.00  &         48.5 &         16.5 &  70.0 &  15.0 &    \textbf{39.0} &  \textbf{65.5} &  17.5 &    64.5 &   76.5 \\
    $\gamma$ = 0.03  &\textbf{62.5} &         20.0 &  74.5 &  15.0 &    26.5 &  63.5 &  15.5 &    71.5 &   79.0 \\
    $\gamma$ = 0.10  &         58.5 &\textbf{27.0} &\textbf{75.0} & \textbf{17.5} &    12.5 &  47.0 &  18.5 &    86.5 &   88.0 \\
    $\gamma$ = 0.30  &         56.0 &         25.5 &  70.5 &   5.0 &     1.5 &  21.5 &  27.0 &    95.0 &   96.5 \\
    $\gamma$ = 1.00  &         39.5 &         16.5 &  65.5 &   1.0 &     0.0 &   2.0 &  \textbf{45.5} &   \textbf{100.0} &  \textbf{100.0} \\
    \bottomrule  
    \end{tabular}
\end{adjustbox}
    \caption{Attack accuracy with different $\gamma$. The magnitude of effective $\gamma$ depends on the reliability of the quality of prior data (see also Tab.~\ref{tab:prior:ssim}).} 
    \label{tab:result:gamma:accuracy}
\end{table}

\begin{table} 
\begin{adjustbox}{width=\columnwidth,center}
    \begin{tabular}{c|ccc|ccc|ccc}
    \toprule
    Dataset & \multicolumn{3}{|c|}{FaceScrub} & \multicolumn{3}{|c|}{LAG} & \multicolumn{3}{|c}{LFW} \\
    \midrule
    Scheme &                  DS-FL & FedGEMS &  FedMD &  DS-FL & FedGEMS &  FedMD &  DS-FL & FedGEMS &  FedMD \\
    \midrule
    $\gamma$ = 0.00  &                  0.457 &   0.359 &  0.532 &  0.163 &   0.286 &  0.354 &  0.298 &   0.457 &  0.445 \\
    $\gamma$ = 0.03  &                  0.468 &   0.408 &  0.560 &  0.163 &   0.313 &  0.357 &  0.313 &   0.522 &  0.476 \\
    $\gamma$ = 0.10  &                  0.487 &   0.467 &  0.590 &  0.170 &   0.338 &  0.376 &  0.320 &   0.554 &  0.525 \\
    $\gamma$ = 0.30  &                  0.528 &   0.570 &  0.648 &  0.184 &   0.362 &  0.397 &  0.340 &   0.617 &  0.623 \\
    $\gamma$ = 1.00  &                  \textbf{0.617} &   \textbf{0.707} &  \textbf{0.744} &  \textbf{0.286} &   \textbf{0.383} &  \textbf{0.405} &  \textbf{0.383} &   \textbf{0.746} &  \textbf{0.769} \\
    \bottomrule
    \end{tabular}
\end{adjustbox}
    \caption{SSIM between private and reconstructed images with different $\gamma$. } 
    \label{tab:result:gamma:ssim}
\end{table}

\begin{table} 
\scriptsize
\centering
\begin{tabular}{ccccc}
\toprule
& \textbf{FaceScrub} & \textbf{LAG} &  \textbf{LFW} \\
\midrule
SSIM & 0.860 & 0.483 & 0.923 \\
\bottomrule
\end{tabular}
\caption{SSIM between the private images and the prior images}
\label{tab:prior:ssim}
\end{table}

\paragraph{Impact of $\tau$}

\begin{figure} 
    \centering
    \includegraphics[width=\linewidth]{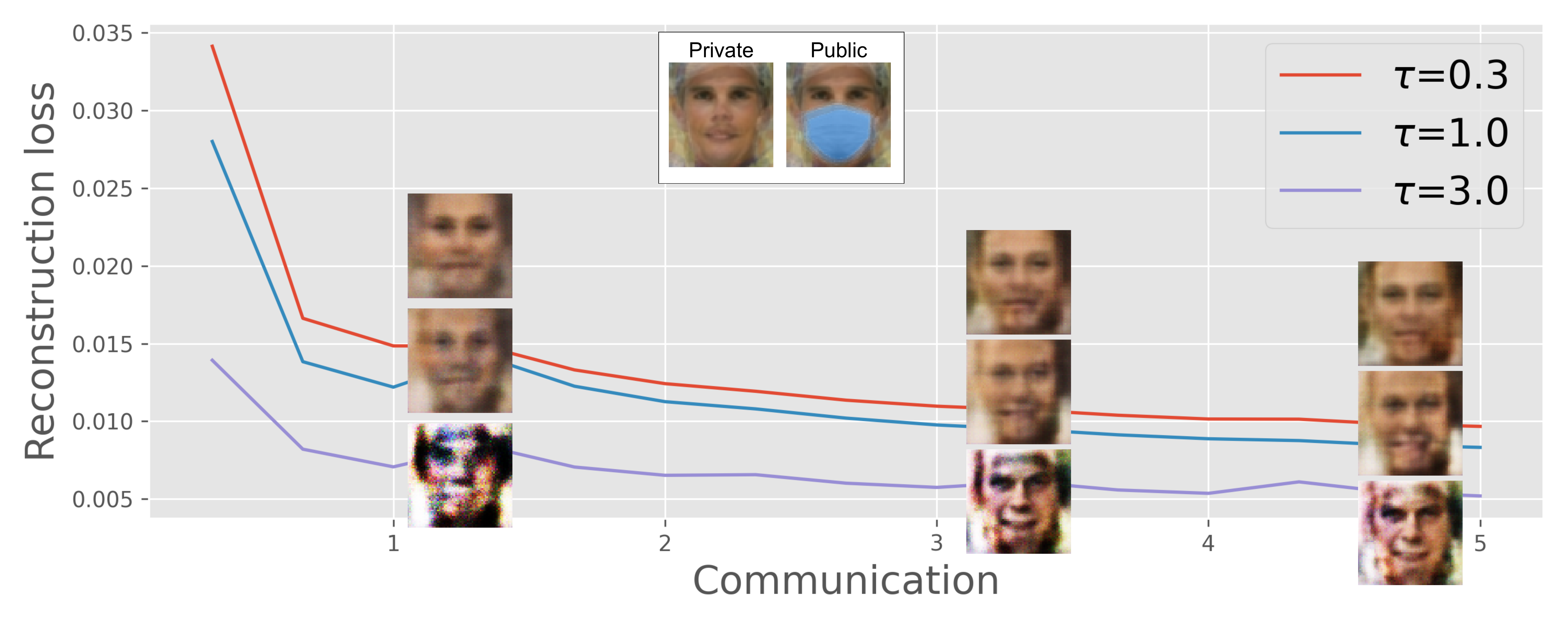}
    \caption{Attack against FedMD in progress on LFW. Increasing $\tau$ makes convergence faster.}
    \label{fig:progeress}
\end{figure}

\begin{figure*}
    \centering
    \includegraphics[height=8.5cm]{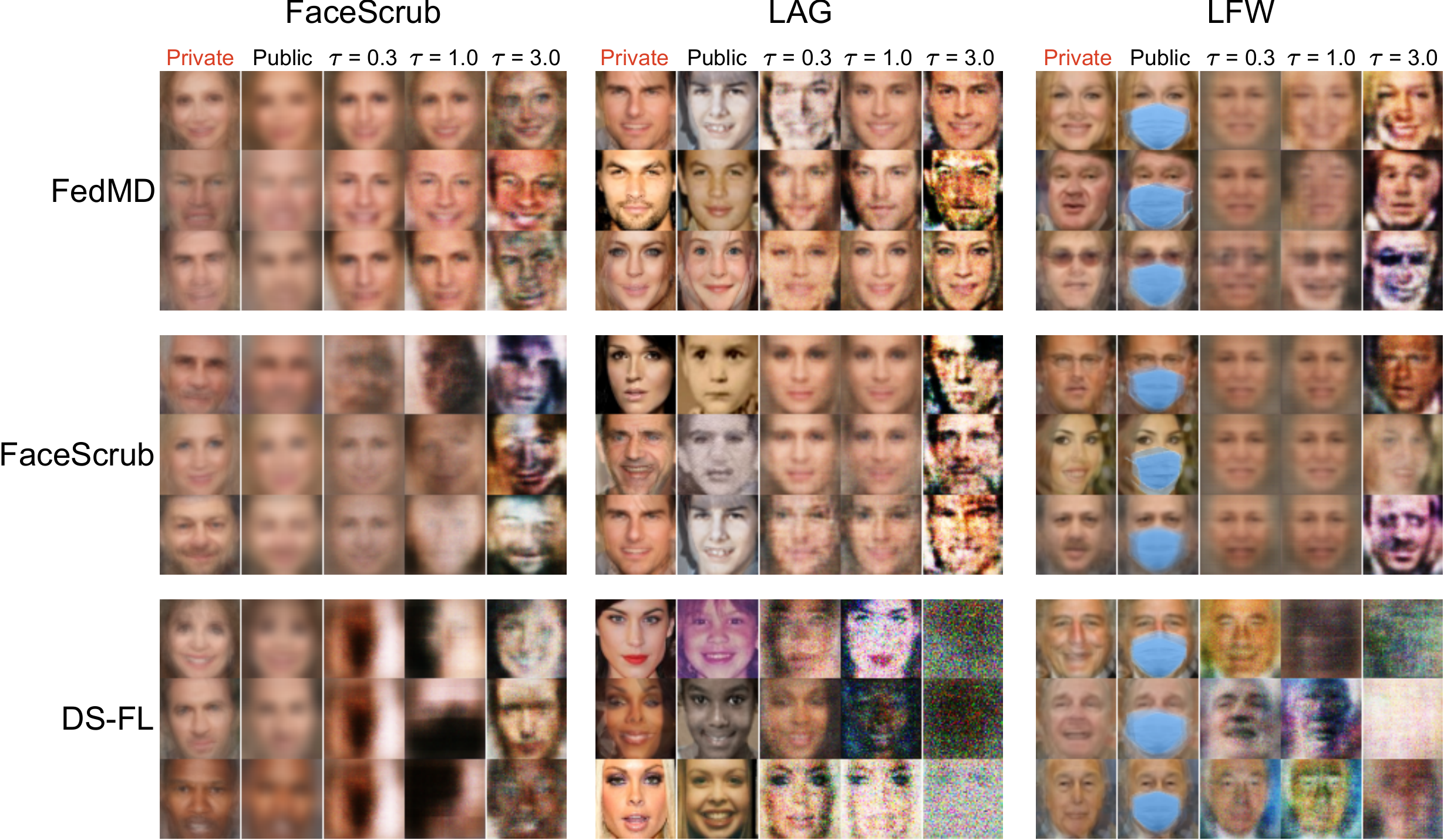}
    \caption{Example of reconstructed images with various $\tau$. Higher $\tau$ helps preserve the unique features of each individual but makes the reconstructed images noisier, especially for FedMD and FedGEMS. The effective $\tau$ is lower in some cases of DS-FL.}
    \label{fig:examples:tau}
\end{figure*}

As stated in Sec.~3.4, $\tau$ controls the trade-off between the quality and the accuracy. Tab.~\ref{tab:result:tau:accuracy} and Tab.~\ref{tab:result:tau:ssim} report the numerical values of attack accuracy and SSIM with different $\tau$, which are summarized in Fig.9. Fig.~\ref{fig:progeress} also shows the reconstruction error and the intermediate recovered images with different $\tau$ against FedMD on LFW dataset, which indicates that larger $\tau$ gives better convergence. Fig.~\ref{fig:examples:tau} shows the reconstructed examples with different temperature $\tau$. We can also observe the same trend in TBI (see Tab.~\ref{tab:result:tau:accuracy:tbi} and Tab.~\ref{tab:result:tau:ssim:tbi}).

\begin{table} 
\begin{adjustbox}{width=\columnwidth,center}
    \begin{tabular}{c|ccc|ccc|ccc}
    \toprule
    Dataset & \multicolumn{3}{|c|}{FaceScrub} & \multicolumn{3}{|c|}{LAG} & \multicolumn{3}{|c}{LFW} \\
    \midrule
    Scheme &                  DS-FL & FedGEMS &  FedMD &  DS-FL & FedGEMS &  FedMD &  DS-FL & FedGEMS &  FedMD \\
    \midrule
    $\tau$ = 0.3 &           9.0 &            0.0 &           38.5 &  \textbf{32.5}&     0.0 &   2.5 &  28.0 &     1.0 &   4.5 \\
    $\tau$ = 1.0 &          16.5 &            0.0 &           64.0 &          31.5 &     1.0 &  14.5 &  \textbf{31.0} &     3.5 &  24.5 \\
    $\tau$ = 3.0 & \textbf{62.5} &           20.0 &  \textbf{74.5} &          15.0 &    26.5 &  63.5 &  15.5 &   71.5 & 
    \textbf{79.0} \\
    $\tau$ = 5.0 &          52.0 &   \textbf{43.5}&  \textbf{87.5} &           3.0 & \textbf{30.0}&  \textbf{66.0} & 14.5 & \textbf{75.0} & 71 \\ 
    \bottomrule  
    \end{tabular}
\end{adjustbox}
    \caption{Attack accuracy with different $\tau$. The higher $\tau$ works when the labeled public dataset is available.} 
    \label{tab:result:tau:accuracy}
\end{table}

\begin{table} 
\begin{adjustbox}{width=\columnwidth,center}
    \begin{tabular}{c|ccc|ccc|ccc}
    \toprule
    Dataset & \multicolumn{3}{|c|}{FaceScrub} & \multicolumn{3}{|c|}{LAG} & \multicolumn{3}{|c}{LFW} \\
    \midrule
    Scheme &                  DS-FL & FedGEMS &  FedMD &  DS-FL & FedGEMS &  FedMD &  DS-FL & FedGEMS &  FedMD \\
    \midrule
    $\tau$ = 0.3 &  \textbf{0.699} &  \textbf{0.846} &  \textbf{0.898} & \textbf{0.325} &          0.533 &  0.543 &  \textbf{0.424} &   0.767 &  \textbf{0.745} \\
    $\tau$ = 1.0 &           0.477 &           0.576 &           0.820 &          0.214 & \textbf{0.554} & \textbf{0.572} &  0.401 &   \textbf{0.771} &  0.740 \\
    $\tau$ = 3.0 &           0.468 &           0.408 &           0.560 &          0.163 &          0.313 &  0.357 &  0.313 &   0.522 &  0.476 \\
    $\tau$ = 5.0 &           0.410 &           0.343 &           0.449 &          0.142 &          0.200 &  0.290 &  0.290 &   0.417 &  0.367 \\ 
    \bottomrule
    \end{tabular}
\end{adjustbox}
    \caption{SSIM between private and reconstructed images with different $\tau$.} 
    \label{tab:result:tau:ssim}
\end{table}

\begin{table} 
\begin{adjustbox}{width=\columnwidth,center}
    \begin{tabular}{c|ccc|ccc|ccc}
    \toprule
    Dataset & \multicolumn{3}{|c|}{FaceScrub} & \multicolumn{3}{|c|}{LAG} & \multicolumn{3}{|c}{LFW} \\
    \midrule
    Scheme &                  DS-FL & FedGEMS &  FedMD &  DS-FL & FedGEMS &  FedMD &  DS-FL & FedGEMS &  FedMD \\
    \midrule
    $\tau$ = 0.3 &       1.0 &     0.0 &   0.0 &   2.5 &     0.0 &   0.0 &   2.0 &     0.0 &   0.0 \\
    $\tau$ = 1.0 &       1.0 &     0.0 &   2.5 &   \textbf{7.5} &     0.0 &   0.0 &   5.5 &     0.0 &   3.0 \\
    $\tau$ = 3.0 &       2.0 &     2.0 &   7.0 &   6.5 &     0.0 &   0.0 &  \textbf{17.5} & \textbf{9.5} & 
    10.0 \\
    $\tau$ = 5.0 & \textbf{3.5} & \textbf{3} & \textbf{8} & 2.5 & 0 & \textbf{1.5} & 13 & 7.5 & \textbf{16} \\
    \bottomrule
    \end{tabular}
\end{adjustbox}
    \caption{Attack accuracy of TBI with different $\tau$. The trend is similar to the results of PIL.} 
    \label{tab:result:tau:accuracy:tbi}
\end{table}

\begin{table} 
\begin{adjustbox}{width=\columnwidth,center}
    \begin{tabular}{c|ccc|ccc|ccc}
    \toprule
    Dataset & \multicolumn{3}{|c|}{FaceScrub} & \multicolumn{3}{|c|}{LAG} & \multicolumn{3}{|c}{LFW} \\
    \midrule
    Scheme &                  DS-FL & FedGEMS &  FedMD &  DS-FL & FedGEMS &  FedMD &  DS-FL & FedGEMS &  FedMD \\
    \midrule
    $\tau$ = 0.3             &  \textbf{0.638} &  \textbf{0.731} &  \textbf{0.525} &     \textbf{0.764} &  \textbf{0.486} &  \textbf{0.799} &     \textbf{0.688} &  \textbf{0.492} &  \textbf{0.801} \\
    $\tau$ = 1.0             &  0.536 &  0.489 &           0.436 &     0.539 &  0.480 &  0.792 &     0.495 &  0.464 &  0.780 \\
    $\tau$ = 3.0             &  0.371 &  0.367 &           0.421 &     0.360 &  0.205 &  0.593 &     0.346 &  0.213 &  0.492 \\
    $\tau$ = 5.0             &  0.375 &  0.358 & 0.343 & 0.312 & 0.144 & 0.444 & 0.405 & 0.151 & 0.384 \\
    \bottomrule
    \end{tabular}
\end{adjustbox}
    \caption{SSIM of TBI between private and reconstructed images with different $\tau$.} 
    \label{tab:result:tau:ssim:tbi}
\end{table}

\paragraph{Impact of Public Dataset Size}

Since PLI relies on public knowledge, we also experiment with a smaller public dataset to test its effect. While the default number of sensitive labels is 200, we set the number of clients to 10 and the number of sensitive labels to 500, which decreases the number of insensitive labels and the amount of public dataset. Tab.~\ref{tab:results:smaller} reports results on FedMD on this smaller public dataset, showing that both accuracy and quality decline. This indicates that a larger public dataset leads to more severe privacy violation.

\begin{table}[!th]
\scriptsize
\begin{tabular}{@{}c|ccc|ccc@{}}
    \toprule
    {} & \multicolumn{3}{c}{SSIM} & \multicolumn{3}{c}{Accuracy} \\
    \midrule
     &              \textbf{FaceScrub} &    \textbf{LAG} &    \textbf{LFW} &    \textbf{FaceScrub} &     \textbf{LAG} &     \textbf{LFW} \\
    \midrule
Default size                & 0.560     & 0.357 & 0.476 & 74.5        & 63.5 & 79.0 \\
Smaller size & 0.251     & 0.295 & 0.370 & 72.6      & 47.4   & 73.4   \\ \bottomrule
\end{tabular}
\caption{Results with the smaller public dataset on FedMD with $K$=10. Smaller public dataset damages attack performance.}
\label{tab:results:smaller}
\end{table}

\section{Information Leakage}
\label{subseq:logit-vs-grad}

We compare logit-based attack with standard gradient-based attack via mutual information (MI). \cite{wang2021improving} finds that we can quantify information leakage between the input and output of a system by MI. We prove that the gradient w.r.t. the model's parameters has higher mutual information between input logits than output. Following Inequal.~\ref{ineq:mi-iq} suggests that gradient can leak more information about the input than the output logit does. 

\begin{prop}
\label{prop:mi-iq}
    Let a neural network contain a biased fully-connected layer as the last layer with a differentiable activation function $y = h(Az + b)$, where $h$ is the activation function; $y \in \mathbb{R}^{N_y}$, $z \in \mathbb{R}^{N_z}$, $A \in \mathbb{R}^{N_y \times N_z}$, $b \in \mathbb{R}^{N_y}$ are the output, input, weight and bias of the last layer, respectively. Then, if $\frac{\partial L}{\partial b}$ is not a zero vector, we have;
    \begin{equation}
    \label{ineq:mi-iq}
        I(x; \frac{\partial L}{\partial A}, \frac{\partial L}{\partial b}) \geq I(x; y),
    \end{equation}
    where $L$ is the differentiable loss function, $x$ is the input data, and $I$ denotes mutual information.
\end{prop} 

\noindent The bellow proof is based on Prop 3.1 in~\cite{geiping2020inverting}.

\begin{proof}
    Since $h$ is differentiable, we have the following equations;
    \begin{align}
        \frac{\partial L}{\partial b_i} &= \frac{\partial L}{\partial y_i} \frac{\partial y_i}{\partial b_i} = \frac{\partial L}{\partial y_i} h^{(1)}(Az+b)_i, \\
        \frac{\partial y_i}{\partial A_i} &= h^{(1)}(Az+b)_i z^{T}
    \end{align}
    , where $i$ is the index of A's row. From the above equations, we can analytically determine $z$ from $\frac{\partial L}{\partial b_i}$ and $\frac{\partial L}{\partial A_i}$ as follows;
    \begin{align}
        z^{T} &= \frac{1}{h^{(1)}(Az+b)} \frac{\partial y_i}{\partial A_i}  \\
              &= \frac{1}{h^{(1)}(Az+b)} \frac{\partial y_i}{\partial L} \frac{\partial L}{\partial A_i} \\
              &= \frac{\partial L}{\partial A_i} / \frac{\partial L}{\partial b_i}
    \end{align} Then, if we think the neural network as a Markov chain $x \rightarrow z \rightarrow y$, the data processing inequality~\cite{cover1999elements} leads to Inequal.~\ref{ineq:mi-iq};
    \begin{align}
        I(x; \frac{\partial L}{\partial A}, \frac{\partial L}{\partial b}) \geq I(x; z) \geq I(x; y)
    \end{align}
\end{proof}

\section{Gradient Inversion Attack}
\label{appendix:sec:gradinv}

\begin{algorithm}[!th]
\caption{Gradient inversion attack}
\label{alg:gbi}
\begin{algorithmic}

\Require The number of communication $T$, the target model $F$, the number of clients $K$, the number of classes $J$, the number of classes of each private dataset $\{J_i\}_{i=1...C}$, the dimension of input $d$.
\Ensure Reconstructed data $\{X'_{i} \in \mathbb{R}^{d \times J_i} \}_{i=1...C}$

\For{$t = 1 \leftarrow T$}
    \For{$i = 1 \leftarrow C$}
        \State The server receives $\nabla W_i$ from client $k$.
        
        \If {$t == 1$}
            \State Infer $Y_i$. 
            \State $X'_{i} \in \mathbb{R}^{d \times J_i} \leftarrow \mathcal{N}(0, 1)$
        \EndIf
        
        \For{$m = 1 \leftarrow M$}
            \State $\nabla W'_{i} \leftarrow \frac{\partial \ell (f(X'_{i}, W_{i}), Y_i)}{\partial W_{i}}$
            \State $X'_{i} \leftarrow X'_{i} - \eta \nabla_{X'_{i}}L_{GB}(X'_{i})$
        \EndFor
    \EndFor
\EndFor
\\
\Return $\{X'_{i}\}_{i=1...C}$

\end{algorithmic}
\end{algorithm}

Although the existing gradient inversion methods focus on reconstructing the exact batch data and labels, our interest is in recovering the class representation of the private training dataset. Then, we view that the received gradient $\nabla W_{i}$ is calculated with $X_{i} \in \mathbb{R}^{J_i \times d}$, where $X_{i}$ represents the class representations of client $k$'s private dataset, $J_i$ is the number of unique classes of the dataset, and $d$ is the dimension of the input data. The attacker can infer the labels used to train the local model from the received gradient with the batch label restoration method proposed in~\cite{yin2021see}. Then, we optimize dummy class representations $X'_{i} \in \mathbb{R}^{J_i \times d}$ with the following cost function;

\begin{align}
    L_{GB}(X'_{i}) = 1 - \frac{\langle \nabla W'_{i}, \nabla W_{i} \rangle}{||\nabla W'_{i}|| ||\nabla W_{i}||} + \gamma \hbox{TV}(X'_{i})
\end{align}

, where $TV$ denotes the total variation and $\gamma$ is its coefficient. This cost function is the same as the one used in~\cite{geiping2020inverting}. Note that unlike our proposed attack against FedMD-like schemes, the attacker must know the number of unique labels in each local dataset in advance. In our experiments, we set $\gamma$ to 0.01, and use Adam optimizer with a learning rate of 0.3.

\begin{table} [!th]
    \centering
    \scriptsize
            \begin{tabular}{ccc}
            \toprule
            \textbf{LFW}    & \textbf{LAG} & \textbf{FaceScrub}    \\ 
            85.5\% & 98.5\% & 95\% \\ \bottomrule
            \end{tabular}
		\caption{Gradient inversion attack against FedAVG ($K=10$). Attack accuracy is higher than logit-based attacks on all datasets.} 
		\label{tab:grad-fedavg}
\end{table}

Tab.~\ref{tab:grad-fedavg} reports the accuracy of gradient inversion attack against FedAVG for 10 clients. Note that this gradient inversion attack does not utilize the prior data based on the public dataset. Across all three datasets, the attack accuracy is higher than PLI without prior data ($\gamma$ = 0.0), which indicates that gradients can potentially leak more private information than PLI without (see Prop.~\ref{prop:mi-iq} in Appendix E). 

\end{document}